\documentclass[lettersize,journal]{IEEEtran}
\usepackage{amsmath,amsfonts}
\usepackage{algorithmic}
\usepackage{algorithm}
\usepackage{array}
\usepackage[caption=false,font=normalsize,labelfont=sf,textfont=sf]{subfig}
\usepackage{textcomp}
\usepackage{stfloats}
\usepackage{url}
\usepackage{verbatim}
\usepackage{graphicx}
\usepackage{cite}
\usepackage{nomencl}
\usepackage{amssymb}
\usepackage{mathtools}
\usepackage{steinmetz}
\usepackage{amsfonts}
\usepackage{amsmath}
\usepackage{graphicx}
\usepackage{multirow}
\usepackage{enumerate}
\usepackage{enumitem}
\usepackage{booktabs}

\newlist{contributions}{enumerate}{1}
\setlist[contributions, 1]
{label=\arabic{contributionsi}., 
leftmargin=\parindent,
rightmargin=10pt
}
\DeclareMathOperator{\diag}{diag}
\DeclareMathOperator{\dv}{d}
\newcommand{\mysum}[2]{\smashoperator[r]{\sum_{#1}} {#2}} 
\newtheorem{remark}{Remark}
\newcommand{\ra}[1]{\renewcommand{\arraystretch}{#1}}

\begin{document}

\title{Parameterized Linear Power Flow for High Fidelity Voltage Solutions in Distribution Systems}

\author{Marija~Markovi\'{c},
        Bri-Mathias~Hodge
\thanks{M. Markovi\'{c} and B.-M. Hodge are with the Department of Electrical, Computer and Energy Engineering and the Renewable and Sustainable Energy Institute at the University of Colorado Boulder, Boulder, CO, 80309 USA. }
\thanks{B.-M. Hodge is also with the National Renewable Energy Laboratory (NREL), Golden, CO 80401 USA.}
\thanks{This work was authored in part by the National Renewable Energy Laboratory (NREL), operated by the Alliance for Sustainable Energy, LLC, for the U.S. Department of Energy (DOE) under Contract No. DE-AC36-08GO28308. The views expressed in the article do not necessarily represent the views of the DOE or the U.S. Government. The U.S. Government retains and the publisher, by accepting the article for publication, acknowledges that the U.S. Government retains a nonexclusive, paid-up, irrevocable, worldwide license to publish or reproduce the published form of this work, or allow others to do so, for U.S. Government purposes.}}




\maketitle

\begin{abstract}
This paper introduces a new model for highly accurate distribution voltage solutions, coined as a parameterized linear power flow model. The proffered model is grounded on a physical model of linear power flow equations, and uses learning-aided parameterization to increase the fidelity of voltage solutions over a wide range of operating points. To this end, the closed-form analytic solution of the parameterization approach is obtained via a Gaussian Process using a deliberately small input sample and without the need for recomputation. The resulting ``self-adjusting" parameter is system-specific and controls how accurate the proposed power flow equations are according to loading conditions. Under a certain value of the resulting parameter, the proposed model can fully recover the linearized formulation of a specialized branch flow model for radial distribution systems, the so-called simplified DistFlow model. Numerical examples are provided to illustrate the effectiveness of the proposed model as well as the improvement in solution accuracy for voltage magnitudes over the simplified DistFlow model and several other linear power flow models, at multiple loading levels. Simulations were carried out on six small- and medium-sized test systems.
\end{abstract}


\section{Introduction}
\IEEEPARstart{T}{he} nonlinearity of the exact AC power flow equations poses difficulties in solving optimization-based problems central to the analysis and operation of power systems. To tackle this challenge, research efforts have developed various approximations and convex relaxations to reduce the computational complexity of power system analysis with nonlinear power flow equations \cite{molzahn2019survey, 6756976, 6507355}. Despite a number of well-established approximate linear models in the existing literature, this is still an active area of research. 

In this paper, we consider the distribution power flow problem under a wide range of changes in the operating conditions. In particular, we present and analyze a new model to linearly characterize the relationship between squared voltage magnitudes and net power injections via kernel-based parameterization. To this end, we use a Gaussian Process (GP), a well-known Bayesian method for non-parametric function estimation \cite{rasmussen2006gaussian}. Note that parameterization here refers to a mathematical function to be estimated from a limited set of operating points. Specifically, this paper leverages GPs to learn a nonlinear function that appears in the derived model, but may be unknown \textit{ab initio} (if no prior knowledge of system states is assumed). This allows the estimated function to be obtained in a closed form using only a small amount of data; as a result, the model preserves a closed-form analytical solution. While the proposed model is novel, GP regression has been used in previous relevant works. For example, GP is utilized as a surrogate model for implicit mapping $y=f^{-1}(x)$ in \cite{9052481,9552521}, where $x$ and $y$ are net power injections and voltages at all buses, respectively. In what follows, we will briefly review existing linear power flow models, discuss the motivation for this work, and summarize the contributions of this paper.

To date, numerous linear models have been proposed (for a comprehensive overview, the interested reader is referred to \cite{molzahn2019survey} and the references therein). The most widely adopted by industry practitioners is the DC power flow (DCPF) model \cite{4956966}. Due to the lack of consideration to the reactive power components, the DCPF model is not suitable for networks with high resistance-to-reactance $\left(r/x\right)$ ratios and where voltage magnitudes and/or reactive branch flows are the primary concerns \cite{4956966, 6981994, 7954975}. Accordingly, various linear models have been developed specifically for the distribution power flow problem, which we roughly classify into three categories: (1) traditional mathematical models, (2) data-driven models, and (3) hybrid models. Needless to say, this categorization is not unique, and the models reviewed below may potentially be classified into several different categories that may even overlap.

Linear models in the first category are derived using (\textit{i}) empirical mathematical approximation, including curve fitting approaches, (\textit{ii}) transformation of state variables (e.g., logarithmic transform), or (\textit{iii}) first-order Taylor series expansion of nonlinear power flow equations. These models are typically based upon certain assumptions (such as the assumption of small voltage angle differences across the branches, and/or the assumption of a flat voltage profile, i.e., $|V|\approx1$), and their accuracy largely depends on a properly selected operating point. A popular example is the lossless formulation of the well-known branch flow model (BFM) for single-phase networks of \cite{19265, 19266, 25627}, the so-called \textit{Simplified DistFlow} model. Its other variants (e.g., multi-phase extensions or generalizations applicable to both radial and meshed topology configurations) can be found, among other works, in \cite{7038399, 7317598, 7244261}. The \textit{Simplified DistFlow} model---henceforth abbreviated \textit{sDistFlow}---gives overestimates of voltage magnitudes \cite[Lemma 12-4]{6756976}. Moreover, its performance weakens with a deviation from the operating point corresponding to lightly loaded networks (i.e., zero-load operating point). Nonetheless, this simplified formulation (or some variant of it) is arguably the most studied model for analysis and operation of distribution networks in the existing literature \cite{7741261, 7317598, 8879622, 8873667, 9466427, 7244261, 8046055}. In fact, multiple models in the first category are derived from (simplified) BFM equations e.g., \cite{7244261, 8046055, 8907467}. While progress has been made in improving the accuracy of the above-mentioned equations, the models of \cite{7244261, 8046055} yield satisfactory results only within a predefined operating region. Using various optimization-based loss parameterizations, \cite{8907467} proposed a model named \textit{Lossy DistFlow}. This model is optimally parameterized using synthetically generated feeders and attains consistently good performance for synthetic test cases at different loading levels, though not for commonly used test cases in the literature.

Models based on the first-order Taylor approximation are reported in \cite{8468081, 7027253, 8332975, 7524028}; these are deemed to be good local linear approximators. Their approximation error, however, increases as the exact operating point moves away from the linearization point. Curve fitting has also been used as a means of linearizing power flow equations (see e.g., \cite{7422165, 8746165}). However, the considered ranges for system states from which the parameters are calculated may noticeably affect the model accuracy; this is due to the fact that the parameters obtained by curve fitting are fixed values \cite{8746165}. 

More recently, the use of data-driven techniques, alone \cite{8289421, 8924608, 9537673} or in addition to physics-based models \cite{9005196, 9626603}, has drawn substantial attention. The former (second category) uses historical operational data to learn mapping rules between observed system input and output, thus circumventing the need for branch parameters and topological knowledge (which is not often available nor reliable in distribution networks). The latter (third category) incorporates knowledge of the physical system into data-driven models and thus may have better performance over a wider range of system operating conditions. Data-driven and hybrid linear power flow models have indeed shown impressive performance but without assurance on the quality of the solution. Nonetheless, their realization relies on critical but difficult-in-practice assumptions about the availability of sufficient historical measurements, especially voltage measurements \cite{8466598}, given the relative paucity of sensing devices and corresponding communications in most distribution systems. 

All reported methodologies, including model-based and data-driven, have strengths in some aspects and weaknesses in others. As aforementioned, conventional (first category) linear models generally exhibit a fixed linearization formula regardless of the loading conditions. Hence, they provide the best approximation accuracy around the linearization point, but suffer from poorer performance with deviation from the assumed operating point. To address this shortcoming and motivated by \cite{8907467}, we propose a new linear power flow model tailored to a wider range of operating points via parameterization. In this paper we particularly highlight the use of GPs for parameterization in linear power flow models. Generally, GP allows greater flexibility over parametric models, such as polynomial regression and neural networks, in that the form of functional relationships it estimates is not predetermined. A GP is particularly data-efficient; it can be trained on small amounts of readily available data \cite{rasmussen2006gaussian}. There are some additional strengths worth noting with this parameterization. While our model is primarily suited to handling steady changes and may be preferable in a slow-changing setting, it could potentially be made suitable to a rapidly time-varying setting as opposed to other conventional models. This, in turn, requires adopting a GP formulation that allows for the inferred function to vary with time. This is especially important for a rapidly changing distribution system landscape with active generation from distributed resources.

It is noteworthy that the formulation detailed here may fall into the third category, though with the advantage that it does not require any historical measured data. Given the differing input data requirements, and thus a lack of clearly defined conditions under which models across different categories could be compared in the existing literature, data-driven models will not be considered here as a basis for comparison. Before proceeding, it is important to point out potential applications in which our model may offer advantages over the less accurate \textit{sDistFlow}, which is routinely used in the literature \cite{7317598, 7244261, 8873667, 9466427}. In general, the proffered model could be applied for distribution system analysis and operations where distribution power flow modeling is needed. It may be of particular interest in applications where trustworthy and fast voltage solutions are critical; notably, timely detection of overvoltages in low-observability distribution grids to comply with power quality standards; or hosting capacity analysis where varying added degrees of distributed generation (DG) and their impact on voltage profiles are studied. Another relevant application arises from the inherently available confidence intervals that GP assigns to each estimate, making probabilistic power flow a practically pertinent extension. This readily available extension requires modification only in the form of the use of input data that accounts for the uncertainty arising from variable net loads (see, for example, \cite{9052481}). 

The contributions of this paper (\textbf{C1}$-$\textbf{C3}) are summarized in the following.
\begin{enumerate}[label=\textbf{C{\arabic*}:}]
\item A new model---coined as a parameterized linear power flow model---has been proposed. The proposed model closely approximates squared voltage magnitudes as a linear function of net power injections, and is characterized by relatively high accuracy in voltage solutions not only at base load, but also at increased loading levels. Although derived independently, this new formulation preserves the mathematical simplicity of the \textit{sDistFlow} equations. By exploiting the tree-like structure of radial distribution networks, a compact matrix-vector formulation of the proposed model is derived for both single-phase and three-phase systems. 
\item A GP is used to optimally parameterize the model herein developed in an \textit{offline} setting. Once parameterized, the model can be used as is. The upshot of the GP-aided parameterization is that it generalizes the proposed model to extended range of valid operating conditions. During operation, point estimates and confidence intervals for voltage magnitudes are obtained in closed-form solution. 
\item Finally, the model is extensively validated using standard test cases for both base load and heavier loading scenarios, thus covering both low-voltage and high-voltage system conditions. Simulations are carried out on six small- and medium-sized test systems. Numerical results corroborate that the voltage solution of our model improves upon voltage solutions of several representative models from the literature.
\end{enumerate}

The remainder of the paper is organized as follows. Section II introduces the distribution system model used throughout this paper. In Section III the proposed model is presented. Section IV reviews GPs and adopts them to parameterize the model. Numerical results are presented in Section V, and the work concludes with Section VI, including future directions. Throughout Sections \ref{Section II} to \ref{Section V}, we restrict our attention to single-phase systems for the sake of exposition.

\section{System Model} \label{Section II}
In this section, we first establish mathematical notation and then describe the adopted network model. We also include a brief overview of (simplified) BFM as it serves as the basis for comparison with the model developed herein. Due to terminology conventions, the terms \textit{bus} (resp. \textit{line} or \textit{branch}) and \textit{node} (resp. \textit{edge}) are used interchangeably hereinafter.

\subsection{Notation} \label{Subsection II-1}
Let $\mathbb{R}$ and $\mathbb{C}$ denote the sets of real and complex numbers, respectively. $\mathcal{R}_{e}\{\cdot\} \in \mathbb{R}$ and $\mathcal{I}_{m}\{\cdot\} \in \mathbb{R}$ respectively return the real and imaginary parts of a complex number, where $j \coloneqq \sqrt{-1}$. Upper-case bold (both upper- and lower-case bold italic) letters are used for matrices (column vectors) e.g., $\textstyle \mathbf{A}$ $\left(\boldsymbol{A}, \boldsymbol{a}\right)$. Non-bold letters are reserved for scalars, e.g., $\boldmath{A, \; a}$. Operators $(\cdot)^{*}$, $(\cdot)^{\top}$, $(\cdot)^{-1}$, $(\cdot)^{-\top}$, and $|\cdot|$ denote complex conjugate, matrix/vector transpose, square matrix inverse, transpose of square matrix inverse, and the absolute value of a number (resp. the component-wise absolute value of a vector/matrix) or the cardinality of a set, respectively. Symbols $\oslash$ and $\odot$ denote the entry-wise (Hadamard) division and multiplication, respectively. For a given vector $\boldsymbol{a} \in \mathbb{R}^{n}$, $\diag\{\boldsymbol{a}\}$ returns a diagonal matrix $\mathcal{D}\left(\boldsymbol{a}\right) \in \mathbb{R}^{n \times n}$ having the entries of $\boldsymbol{a}$ on its diagonal, $\|\boldsymbol{a}\|_{\infty} \coloneqq \max\left(|a_{1}|,...,|a_{n}|\right)$, and $\|\boldsymbol{a}\|_{1} \coloneqq \sum_{i=1}^{n}|a_{i}|$. A matrix operator $\dv\left(\cdot\right)$ takes a square matrix as an input and returns its diagonal elements as a column vector. $\boldsymbol{1}_{n}\in \mathbb{R}^{n}$, $\boldsymbol{0}_{n}\in \mathbb{R}^{n}$, and $\mathbf{I}_{n} \in \mathbb{R}^{n \times n}$ are respectively a unity column vector, a zero column vector, and the identity matrix, which will occasionally be abbreviated as $\boldsymbol{1}$, $\boldsymbol{0}$, and $\mathbf{I}$. With a slight abuse of the notation, $\boldsymbol{x}$ and $\mathbf{x}$ will be used for the branch reactance vector and the input vector used in the Section \ref{Section IV}, respectively; additionally, $\ell$ w/ and w/o subscript will denote the squared current magnitude along the corresponding branch and the generic distribution line, respectively, which allows us to be consistent with the literature, as well as not introducing additional symbols.
 
\subsection{Modeling of Radial Distribution Network} \label{Subsection II-2}
Consider a radial distribution system represented by a \textit{directed rooted tree} graph $\mathcal{G}(\mathcal{N}\cup \{0\},\mathcal{L})$ comprising $|\mathcal{L}|$ lines collected in the set $\mathcal{L}$, $\mathcal{L}\coloneqq\{1,...,\ell\}$ and $|\mathcal{N}| + 1$ buses collected in the set $\mathcal{N} \cup \{0\}$, $\mathcal{N}\coloneqq\{1,...,n\}$, where bus indexed 0 is the substation bus i.e., the root of the tree $\mathcal{G}$. Each edge $\ell=(i,j) \in \mathcal{L}$ is associated with an ordered pair of nodes it connects $i,j \in \mathcal{N}$, which are denoted as the \textit{sending-end} and \textit{receiving-end} nodes, respectively. Let $\mathcal{N}_{\mathfrak{D}}(i) \subseteq \mathcal{N}$ denote the set of all nodes located downstream of node $i$ including node $i$ itself. Without loss of generality, it is assumed that the complex voltage of the root node is known and fixed, and that each non-root node hosts a load and possibly also an inverter-interfaced DG. The topology of such a system is uniquely determined by the edge-to-node incidence matrix $\bar{\mathbf{M}}=[\boldsymbol{m}_{0} \; \mathbf{M}] \in \mathbb{R}^{\ell \times \left(n + 1\right)}$ defined componentwise as:
\begin{equation*} 
    \bar{\mathbf{M}} = 
     \begin{cases}
      1,&{\text{if line}}\ \ell \in \mathcal{L}\ {\text{leaves node}}\ i \in \mathcal{N} \cup \{0\} \\
      -1,&{\text{if line}}\ \ell \in \mathcal{L}\ {\text{enters node}}\ i \in \mathcal{N} \\
      {0,}&{\text{otherwise,}} 
    \end{cases} 
\end{equation*}
where $\mathbf{M} \in \mathbb{R}^{\ell \times n}$ is the reduced incidence matrix of $\mathcal{G}$ that results from removing the first column of $\bar{\mathbf{M}}$ corresponding to the root node (i.e., $\boldsymbol{m}_{0})$ \cite{7317598}. Under the assumption of radial network topology, $\mathbf{M}$ is nonsingular with the following property $\mathbf{M}^{-1}\boldsymbol{m}_{0}=-\mathbf{1}_{n}$ \cite{7317598}. 

\begin{figure}[!t]
\centering
\includegraphics[height=3.5cm]{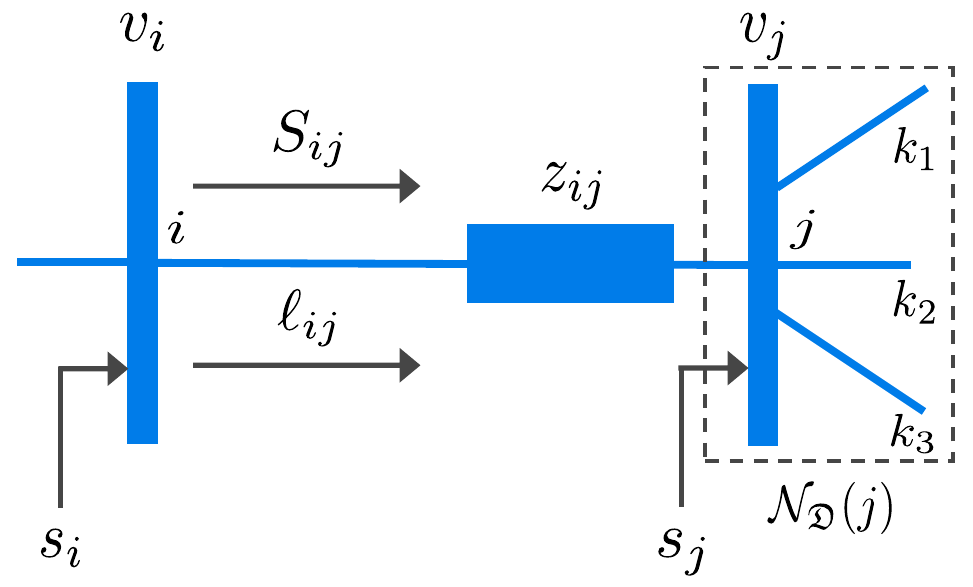}
\caption{A two-bus system representing a distribution line $(i,j) \in \mathcal{L}$ between sending-end $i$ and receiving-end $j$ with respective electrical quantities.}
\label{1}
\end{figure}
 
Some of the notations introduced below are depicted in Fig.~1. For each branch $(i,j) \in \mathcal{L}$, let $z_{ij} = r_{ij} +jx_{ij}$ be its complex impedance and define $\boldsymbol{r}=[r_{1},...,r_{n}]^{\top} \in \mathbb{R}^{n}, \; \boldsymbol{x}=[x_{1},...,x_{n}]^{\top} \in \mathbb{R}^{n}$; let $S_{ij} = P_{ij} +jQ_{ij}$ denote the sending-end complex branch power flow and define $\boldsymbol{P}=[P_{1},...,P_{n}]^{\top} \in \mathbb{R}^{n}, \; \boldsymbol{Q}=[Q_{1},...,Q_{n}]^{\top} \in \mathbb{R}^{n}$; and let $\ell_{ij} \coloneqq |I_{ij}|^2$ denote the squared current magnitude and define $\boldsymbol{\ell}=[\ell_{1},...,\ell_{n}]^{\top} \in \mathbb{R}^{n}$. For each node $i \in \mathcal{N}$, let $V_{i} = |V_{i}| \angle \delta_{i} \in \mathbb{C}$ be a line-to-ground voltage phasor, where $|V_{i}| \in \mathbb{R}$ is a voltage magnitude and $\delta_{i} \in \mathbb{R}$ is a voltage angle with respect to an arbitrary system reference, typically the root node; and let $s_{i} = p_{i} +jq_{i}$ be the net complex power injection. Define further the (squared) voltage magnitude vectors $\left(\boldsymbol{v}\coloneqq[|V_{1}|^2,...,|V_{n}|^2]^{\top} \in \mathbb{R}^{n}\right) \;\boldsymbol{V}\coloneqq[|V_{1}|,...,|V_{n}|]^{\top} \in \mathbb{R}^{n}$ with corresponding root-node variables discarded (i.e., $v_{0}, |V_{0}|$). Also, let $\boldsymbol{p} = [p_{1},...,p_{n}]^{\top} \in \mathbb{R}^{n}, \; \boldsymbol{q} = [q_{1},...,q_{n}]^{\top} \in \mathbb{R}^{n}$ respectively collect active and reactive net power injections at all non-root nodes, where $\boldsymbol{p}$ (resp. $\boldsymbol{q}$) is decomposed into power generation and consumption components, i.e., $\boldsymbol{p}=\boldsymbol{p}^{g}-\boldsymbol{p}^{d}$ (resp. $\boldsymbol{q}=\boldsymbol{q}^{g}-\boldsymbol{q}^{d}$). The units of all variables involved are \textit{per unit} (p.u.) unless otherwise noted.

The system defined in this way can be modeled by a single-phase BFM, whose compact form \eqref{eq01:main} has been revised below.
\begin{subequations}\label{eq01:main}
\begin{gather}
\small\boldsymbol{p} = \mathbf{M}^{\top}\boldsymbol{P} + \mathcal{D}\left(\boldsymbol{r}\right)\boldsymbol{\ell} \label{eq01:a}\\ 
\small\boldsymbol{q} = \mathbf{M}^{\top}\boldsymbol{Q} + \mathcal{D}\left(\boldsymbol{x}\right)\boldsymbol{\ell} \label{eq01:b}\\
\small\boldsymbol{v} = v_{0}\mathbf{1}_{n}\!+\!2\mathbf{M}^{-1}\!\left(\mathcal{D}\left(\boldsymbol{r}\right)\!\boldsymbol{P}+\mathcal{D}\left(\boldsymbol{x}\right)\!\boldsymbol{Q}\right)\!-\!\mathbf{M}^{-1}\!\left(\mathcal{D}^{2}\!\left(\boldsymbol{r}\right)+\mathcal{D}^{2}\!\left(\boldsymbol{x}\right)\right)\!\boldsymbol{\ell} \label{eq01:c}
\end{gather}
\end{subequations}
Solving for $\boldsymbol{P}$ and $\boldsymbol{Q}$ respectively from \eqref{eq01:a} and \eqref{eq01:b} and substituting into \eqref{eq01:c} yield a nonlinear equation \eqref{eq02:main} that provides an exact voltage solution.
\begin{multline}\label{eq02:main}
\boldsymbol{v} = v_{0}\mathbf{1}_{n}-\mathbf{M}^{-1}\!\left(\mathcal{D}^{2}\left(\boldsymbol{r}\right)+\mathcal{D}^{2}\left(\boldsymbol{x}\right) \right)\boldsymbol{\ell} + 2\mathbf{M}^{-1}\times \\ \shoveleft{\Bigl(\mathcal{D}\!\left(\boldsymbol{r}\right)\mathbf{M}^{-\!\top}\!\!\left(\boldsymbol{p}-\mathcal{D}\!\left(\boldsymbol{r}\right)\boldsymbol{\ell}\right)+\mathcal{D}\!\left(\boldsymbol{x}\right)\mathbf{M}^{-\!\top}\!\!\left(\boldsymbol{q}-\mathcal{D}\!\left(\boldsymbol{x}\right)\boldsymbol{\ell}\right)\Bigr)} 
\end{multline}
The simplified BFM is derived from \eqref{eq01:main} by omitting terms related to branch losses. Similarly, the exclusion of these terms from \eqref{eq02:main} gives an approximate voltage solution; $\boldsymbol{\tilde{v}}$ hereinafter.

\section{Parameterized Linear Power Flow Model} \label{Section III}
In this section, the derivation and compact formulation of the new linear power flow model are presented. The correlation between the proposed model and the simplified BFM of \cite{19265, 19266, 25627} is also discussed. For simplicity in exposition, the model is outlined for a single-phase system. An extension to three-phase systems can be found in the Appendix B. 

\subsection{Model Derivation} \label{Subsection III-1}
Consider a line segment of the radial distribution network between any two adjacent nodes depicted in Fig.~1. The following derivation holds $\forall\ell=(i,j) \in \mathcal{L}$. As illustrated in Fig.~2, both line resistance and reactance contribute to the corresponding voltage drop that can be expressed as:
\begin{align} \label{eq0:main}
\Delta V_{ij} &= |V_{i}| \angle \delta_{i} - |V_{j}| \angle \delta_{j} = z_{ij}\left(\frac{S_{ij}}{|V_{i}|\angle\delta_{i}}\right)^{*}    
\end{align}
from which its rectangular components can be obtained:
\begin{subequations}\label{eq1:main}
\begin{align}
\Delta V_{ij}^{re} &= |V_{i}| - |V_{j}|\cos(\delta_{ij}) = \frac{\mathcal{R}_{e}\{z_{ij}S_{ij}^{*}\}}{|V_{i}|} \label{eq1:b} \\ 
\Delta V_{ij}^{im} &= |V_{j}|\sin(\delta_{ij}) = \frac{\mathcal{I}_{m}\{z_{ij}S_{ij}^{*}\}}{|V_{i}|} \label{eq1:c}
\end{align}
\end{subequations}

In a distribution system, voltage angles typically vary within relatively narrow limits, and angular differences across lines are generally very small \cite{7422165}. For mathematically close voltage angles between connected nodes, the \textit{small angle assumption} yields the following approximations of their respective trigonometric terms, $\cos{\delta_{ij}} \approx 1 - \frac{1}{2}\delta_{ij}^2 \approx 1$ $\left(\delta_{ij}^2 << \delta_{ij}\right)$ and $\sin{\delta_{ij}} \approx \delta_{ij}, \; \forall \left(i,j\right) \in \mathcal{L}$. Using these simplified trigonometric terms, \eqref{eq1:main} can be solved for squared voltage magnitudes of the sending- and receiving-end nodes, respectively:
\begin{subequations}\label{eq2:main}
\begin{align}
v_{i} &= \frac{\mathcal{I}_{m}\{z_{ij}S_{ij}^{*}\}}{\delta_{ij}} + \mathcal{R}_{e}\{z_{ij}S_{ij}^{*}\} \label{eq2:a} \\ 
v_{j} &= \frac{\mathcal{I}_{m}^{2}\{z_{ij}S_{ij}^{*}\}}{\delta_{ij}^{2}\mathcal{R}_{e}\{z_{ij}S_{ij}^{*}\} + \delta_{ij}\mathcal{I}_{m}\{z_{ij}S_{ij}^{*}\}} \label{eq2:b}
\end{align}
\end{subequations}
Combining \eqref{eq2:a} and \eqref{eq2:b} gives
\begin{equation}\label{eq3}
v_{i} - v_{j}  = \frac{\mathcal{I}_{m}\{z_{ij}S_{ij}^{*}\}}{\delta_{ij}} \left(1-\lambda_{ij}\right) \\ + \mathcal{R}_{e}\{z_{ij}S_{ij}^{*}\},   
\end{equation}
which can be further rewritten into a more compact form
\begin{equation} \label{eq3:main}
v_{i} - v_{j}  = \left(1+\lambda_{ij}\right)\left(r_{ij}P_{ij} + x_{ij}Q_{ij}\right),
\end{equation}
where $\lambda_{ij}$ is introduced for exposition simplicity and is given by:
\begin{equation} \label{eq4}
\lambda_{ij} = \left(1 + \delta_{ij}\frac{\Delta V_{ij}^{re}}{\Delta V_{ij}^{im}}\right)^{-1}
\end{equation}
\begin{remark}
Note the difference between \eqref{eq3:main} and the \textit{sDistFlow} equations, according to which \cite{19265, 19266, 25627}:
\begin{equation} \label{eq:sdistflow}
\tilde{v}_{i} - \tilde{v}_{j}  = 2\left(r_{ij}P_{ij} + x_{ij}Q_{ij}\right)
\end{equation}
Equation \eqref{eq3:main} can be considered a special case of \eqref{eq:sdistflow} if $\lambda_{ij}$ is homogeneous for all line sections and is a unity scalar. That is, \eqref{eq3:main} reduces to \eqref{eq:sdistflow} when $\lambda_{ij}=1, \; \forall (i,j) \in \mathcal{L}$. 
\end{remark}

\begin{figure}[!t]
\centering
\includegraphics[height=2cm]{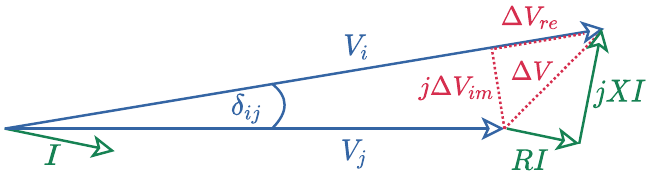}
\caption{Voltage phasor diagram in the system from Fig.~1. The voltage drop along the branch and its rectangular components are denoted by $\Delta V_{ij} \in \mathbb{C}$, $\Delta V_{ij}^{re} = \mathcal{R}_{e}\{\Delta V_{ij} \} \in \mathbb{R}$, $\Delta V_{ij}^{im} = \mathcal{I}_{m}\{\Delta V_{ij} \} \in \mathbb{R}$, respectively.}
\label{2}
\end{figure}

The expression relating branch power flows to squared voltage magnitudes, given by \eqref{eq3:main}, is nonlinear due to: (\textit{i}) $\lambda_{ij}$, and (\textit{ii}) $P_{ij}, Q_{ij}$. To ensure linearity of the model, as described in the following, the second term in \eqref{eq4} is replaced with $\alpha_{ij}$ which simplifies to
\begin{equation} \label{eq4:main}
\lambda_{ij} = \left(1 + \alpha_{ij}\right)^{-1},
\end{equation}
where it can be shown that $\alpha_{ij}$ characterizes the change in voltage magnitude between adjacent nodes relative to the nominal value (i.e., 1 p.u.); see Appendix A for more details. Using a binomial expansion, \eqref{eq4:main} can be approximated with great fidelity to:
\begin{equation} \label{eq4:mod1}
\lambda_{ij} \approx 1 - \alpha_{ij}, \; \text{for} \; |\alpha_{ij}| \ll 1
\end{equation}
which introduces a negligibly small absolute error on the order of $10^{-4}$ on average for the six test systems used to evaluate the model as depicted in Fig~3. 

To derive the relationship between squared voltage magnitudes and nodal power injections from \eqref{eq3:main}, assuming the latter are known variables, we can approximate line power flows by summing all downstream power injections as done in \cite{8907467}. This together with \eqref{eq4:mod1} still results in a nonlinear relationship---due to the unknown term\footnote{The subscript $ij$ is suppressed in notation for brevity under the understanding that technical arguments apply to each of the branch-specific functions.} $\alpha$ (referred to as the \textit{voltage sensitivity} function)---which can be linearized by replacing $\alpha$ with a properly chosen approximate value $\hat{\alpha}$, which is central to our analysis. Accordingly, achieving high approximation accuracy requires a well-approximated $\hat{\alpha}$, which in turn leads to a plausible voltage estimate $\hat{v}$. In this paper, a GP is pursued to approximate $\alpha$ resulting in a closed-form approximation $\hat{\alpha}$ as explained shortly. 

\begin{figure}[!t]
\centering
\includegraphics[height=5cm]{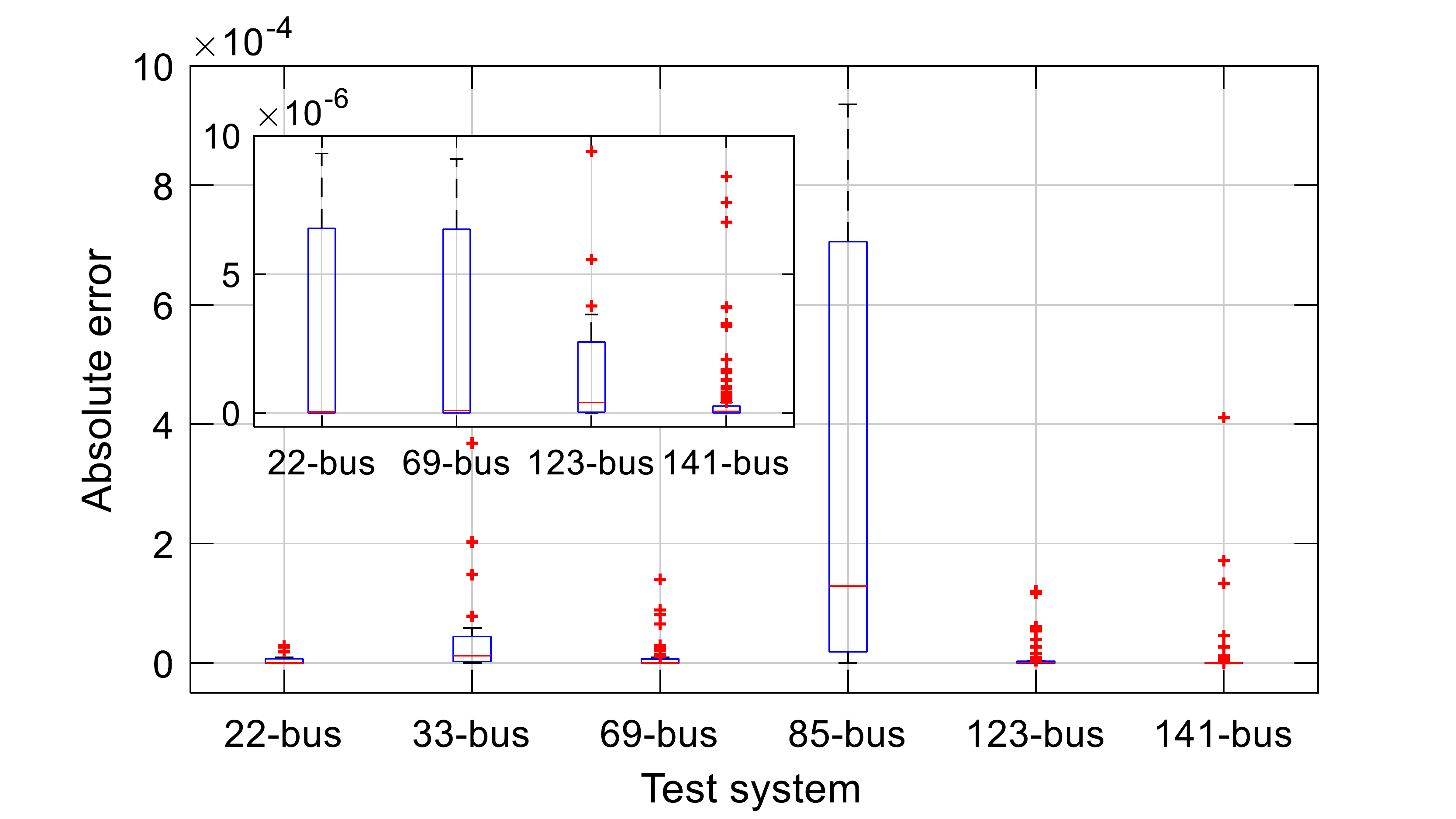}
\caption{Boxplot of absolute errors over $|\mathcal{L}|$ approximations of $\lambda_{\ell} \; \left(\forall \ell \in \mathcal{L}\right)$ for each distribution test system. Absolute errors are calculated as $|\lambda_{\ell}-\hat{\lambda}_{\ell}|$, where $\lambda_{\ell}$ is defined in (\ref{eq4:main}) and $\hat{\lambda}_{\ell}$ is its binomial approximation (\ref{eq4:mod1}). Median error values (horizontal red lines) in the appearance order of the test systems are $4.6 \cdot 10^{-8}$, $1.2 \cdot 10^{-5}$, $9.9 \cdot 10^{-8}$, $1.3 \cdot 10^{-4}$, $3.8 \cdot 10^{-7}$, $6.4 \cdot 10^{-8}$, respectively.} 
\label{3}
\end{figure}

Given the above, \eqref{eq3:main} can be linearly approximated as:
\begin{align} \label{eq5:main} 
\hat{v}_{i} - \hat{v}_{j}  = \left(2-\hat{\alpha}_{ij}\right)\!\bigg(r_{ij}\!\!\!\!\mysum{k \in \mathcal{N}_{\mathfrak{D}}(j)}{\left(-p_{k}\right)} + x_{ij}\!\!\!\!\mysum{k \in \mathcal{N}_{\mathfrak{D}}(j)}{\left(-q_{k}\right)}\bigg)
\end{align}
Per \eqref{eq5:main}, hereinafter the parameterized linear power flow (PLPF) model, squared voltages are approximately affine functions of nodal power injections. It is worth noting that the model is conditionally linear by parameterization, hence the name of the model. Mathematically, the derived formulation preserves the simplicity of the simplified BFM (see \textit{Remark 1}). After solving \eqref{eq5:main}, voltage magnitudes are obtained directly, whereas the voltage angles can be subsequently recovered (see \cite{6507355} for details). 
\begin{remark}
Recall, \eqref{eq5:main} was derived considering only the improvement of the voltage solution accuracy. Despite the approximation of branch power flows by summing all downstream power injections, the model is very accurate. In fact, its performance is determined by $\hat{\alpha}$ which implicitly includes branch losses, as shown next. Importantly, pertinent approaches from the literature that more accurately model branch power flows can be incorporated in \eqref{eq3:main} (e.g., the loss factors proposed in \cite{7524028}), however, this falls outside the scope of the current work. 
\end{remark}

\subsection{Vector-Matrix Model Formulation} \label{Subsection III-2}
While we omit the details (see, for example, \cite{8907467}), a compact vector-matrix form of \eqref{eq5:main} can be written as:
\begin{equation}\label{eq6:main} 
\hat{\boldsymbol{v}} = v_{0}\mathbf{1}_{n} + \mathbf{M}^{-1}\mathbf{\hat{\Lambda}}\left(\mathcal{D}\left(\boldsymbol{r}\right)\mathbf{M}^{-\top}\boldsymbol{p} + \mathcal{D}\left(\boldsymbol{x}\right)\mathbf{M}^{-\top}\boldsymbol{q}\right),
\end{equation}
where
\begin{equation*}
\mathbf{\hat{\Lambda}} = 2\mathbf{I}_{n} - \mathcal{D}\left(\boldsymbol{\hat{\alpha}}\right).    
\end{equation*}
The resulting model \eqref{eq6:main} is computationally very light as it requires only elementary vector-matrix multiplications (after $\boldsymbol{\hat{\alpha}}$ is inferred \textit{offline} via GP-aided parameterization). Before proceeding, we first recast \eqref{eq6:main} into the form
\begin{multline}\label{eq6:mod} 
\small 
\hat{\boldsymbol{v}} = v_{0}\mathbf{1}_{n} + 2\left(\mathbf{M}^{-1}\mathcal{D}\left(\boldsymbol{r}\right)\mathbf{M}^{-\top}\boldsymbol{p} + \mathbf{M}^{-1}\mathcal{D}\left(\boldsymbol{x}\right)\mathbf{M}^{-\top}\boldsymbol{q} \right) \\ \small -\mathbf{M}^{-1}\mathcal{D}\left(\boldsymbol{\hat{\alpha}}\right)\left(\mathcal{D}\left(\boldsymbol{r}\right)\mathbf{M}^{-\top}\boldsymbol{p} + \mathcal{D}\left(\boldsymbol{x}\right)\mathbf{M}^{-\top}\boldsymbol{q}\right),
\end{multline}
from which it is clear that the first two terms are the voltage solution to the \textit{sDistFlow} equations. The third term in \eqref{eq6:mod} is novel and depends on $\boldsymbol{\hat{\alpha}}$, which we refer to as the \textit{voltage sensitivity} vector or, abbreviated, voltage sensitivity. Taken together, \eqref{eq6:mod} is a voltage solution of PLPF. As previously shown, in the case of $\boldsymbol{\hat{\alpha}} = \boldsymbol{0}$ (equivalently, $\mathbf{\hat{\Lambda}} = \boldsymbol{1}$), it follows that \textit{sDistFlow} and PLPF are equivalent. 

Having established the relationship between the proposed model and simplified BFM, we now find it useful to introduce an exact nonlinear expression for voltage sensitivity, given in compact form by
\begin{multline}\label{eq7:main}
\small
\boldsymbol{\alpha} = d\Bigl[ \Bigl(2\left(\mathcal{D}\left(\boldsymbol{r}\right)\mathbf{M}^{-\top}\mathcal{D}\left(\boldsymbol{r}\right) + \mathcal{D}\left(\boldsymbol{x}\right)\mathbf{M}^{-\top}\mathcal{D}\left(\boldsymbol{x}\right)\right) \\ \small  \shoveleft{-\Bigl(\mathcal{D}^2\!\left(\boldsymbol{r}\right) + \mathcal{D}^2\!\left(\boldsymbol{x}\right)\Bigr)\Bigr)\boldsymbol{\ell} \oslash\!\left(\mathcal{D}\left(\boldsymbol{r}\right)\!\mathbf{M}^{-\top}\!\boldsymbol{p} + \mathcal{D}\left(\boldsymbol{x}\right)\!\mathbf{M}^{-\top}\!\boldsymbol{q}\right)\Bigr]}
\end{multline}
as may be verified by equalizing the difference between \eqref{eq02:main} and \eqref{eq6:main} to zero. In \eqref{eq7:main}, $\boldsymbol{\ell}$ can take exact or approximate form, the latter of which results in
\begin{equation*}\label{eq8:main}
\boldsymbol{\ell} \approx \mathcal{D}\left(\boldsymbol{r}^2+\boldsymbol{x}^2\right)^{-1}\mathcal{D}\left(\mathbf{M}\left(\boldsymbol{V}-|V_{0}|\mathbf{1}\right)\right)\mathbf{M}\left(\boldsymbol{V}-|V_{0}|\mathbf{1}\right).
\end{equation*}
An accurate model (i.e., mathematical representation) of voltage sensitivity defined in \eqref{eq7:main} allows us to estimate $\boldsymbol{\hat{\alpha}}$ via GP using a sequence of power flow solutions, as described in more detail in Section \ref{Section IV}. In this paper, we assume that the standard GP is a good model for the voltage sensitivity, which is corroborated by the numerical results. 
\begin{remark}
Recall that $\tilde{\boldsymbol{v}} \geq \boldsymbol{v}$ \cite[Lemma 12-4]{6756976}, where $\boldsymbol{v}$ and $\tilde{\boldsymbol{v}}$ are exact and approximate solutions of the respective BFMs \cite{19265, 19266, 25627}. Then based on \eqref{eq6:mod}, $\hat{\boldsymbol{v}} \leq \tilde{\boldsymbol{v}}$ is required to obtain close-to-accurate voltage estimates via \eqref{eq6:main}, a condition that is met when $|\boldsymbol{\hat{\alpha}}| \ll \boldsymbol{1}$\footnote{Between $|\boldsymbol{\hat{\alpha}}| \leq \boldsymbol{1}$ and $|\boldsymbol{\hat{\alpha}}| \ll \boldsymbol{1}$ from \eqref{eq4:mod1}, the latter is a stronger condition.}, where a positive sign is taken if the nodal power is injected into the grid, and a negative sign otherwise.
\end{remark}

\subsection{Solution Approaches} \label{Subsection III-3}
There are two possible approaches to voltage solution calculation based on the proposed model. The first is known in the literature as \textit{cold-start} method, which directly solves the PLPF without any knowledge of the present system state. The latter is the so-called \textit{warm-start} method, which relies on an AC base-point solution (specifically, voltage magnitudes). Namely, from \eqref{eq7:main}, $\boldsymbol{\alpha}$ is a function of voltage magnitudes (unknown state variables) in addition to nodal power injections (known system inputs), which implies that it cannot be explicitly calculated using system input alone. Instead, it can be determined once the base-point voltage solution is obtained, which in turn increases the complexity of the model. Alternatively, voltage sensitivity (and thus the developed model) can be initialized with a fixed-point value pertinent to the condition from \textit{Remark 3} (e.g., $|\hat{\alpha}_{\ell}| = 10^{-3}, \; \forall \ell \in \mathcal{L}$), and then iteratively updated by taking an appropriately sized step in a direction tangential to the solution trajectory, which is also computationally impractical. 

In this work, we opt for the first approach that does not require iterations. To \textit{cold-start} the PLPF model, we use the GP's posterior mean realized \textit{offline}, as explained next.

\section{Parameterization via Gaussian Processes} \label{Section IV} 
Hereby, we first briefly revisit GP regression, and then detail the aforesaid GP-based parameterization.

\subsection{Gaussian Processes Revisited} \label{Subsection IV-1}
A GP is a stochastic process $g(\mathbf{x})$ that is fully specified by its mean function $m(\mathbf{x})$ and covariance (kernel) function $k(\mathbf{x},\mathbf{x'})$ \cite{rasmussen2006gaussian}. That is, for any input points $\mathbf{x},\mathbf{x'}$, we can write
\begin{gather*} \label{eq01:GP}
g(\mathbf{x})  \sim \mathcal{GP}\left(m(\mathbf{x}), k(\mathbf{x},\mathbf{x'})\right)\\ 
m(\mathbf{x}) = \mathbb{E}[g(\mathbf{x})] \\
k(\mathbf{x},\mathbf{x'}) = \mathbb{E}[(g(\mathbf{x}) - m(\mathbf{x}))(g(\mathbf{x'}) - m(\mathbf{x'}))] 
\end{gather*}
where $\mathbb{E}$ is an expectation operator. Very often, the mean function $m(\cdot)$ is assumed zero. The choice of $k(\cdot,\cdot)$ is crucial because it directly encodes prior assumptions about the true underlying function $g(\cdot)$ \cite{rasmussen2006gaussian}. One very popular choice for learning processes that are known to be smooth is the squared exponential (SE) kernel, also known as the radial basis function, which is defined as:
\begin{equation} \label{eq02:GP}
k_{SE}(\mathbf{x},\mathbf{x'}) = \sigma_{g}^2exp\Big[-\frac{1}{2l^2}\left(\mathbf{x}-\mathbf{x'}\right)^{\top}\left(\mathbf{x}-\mathbf{x'}\right)\Big]  
\end{equation}
where $\sigma_{g}^2$ and $l$ are called signal variance and characteristic length-scale, respectively \cite{rasmussen2006gaussian}. The two hyperparameters ($\sigma_{g}^2$ and $l$) can be efficiently inferred from data  using gradient-based optimization, for example, the maximum likelihood estimator (for more details, please refer to \cite{gelman2013bayesian}). Once $m(\cdot)$ and $k(\cdot,\cdot)$ are chosen, GPs are used to draw \textit{a priori} as well as posterior function values conditioned upon previous observations \cite{rasmussen2006gaussian,gelman2013bayesian}. 

\subsection{Proposed Parameterization} \label{Subsection IV-2}
The exact model of voltage sensitivity introduced by \eqref{eq7:main} can be formalized, with an optional noise term, as:
\begin{equation} \label{eq1:GP}
\boldsymbol{\alpha} = \mathbf{g}(\boldsymbol{s}) + \boldsymbol{\epsilon}, 
\end{equation}
where $\mathbf{g}$ is a nonlinear function modeling $\boldsymbol{\hat{\alpha}}$ that we want to learn via GP regression, $\boldsymbol{s}$ is the vector of net complex power injections at all non-root nodes, and $\boldsymbol{\epsilon}$ is a vector collecting $\epsilon_{\ell} \sim \mathcal{N}\left(0,\sigma_{\epsilon}^2\right)$ i.e., \textit{i.i.d.} Gaussian noise with zero mean and variance $\sigma_{\epsilon}^2$. In the standard \textit{input-output mapping} notation, \eqref{eq1:GP} can be described by a non-parametric model
\begin{equation*} 
\mathbf{y} = \mathbf{g}(\mathbf{X}) + \boldsymbol{\epsilon}
\end{equation*}
where $\mathbf{X}$ and $\mathbf{y}$ are the input matrix and the output vector, respectively, which technically constitute a training data set $\mathcal{D}$. With this reformulation in place, let us now model $\mathbf{g}\left(\cdot\right)$ as a zero-mean GP with covariance defined in \eqref{eq02:GP} (which is advocated in many settings in \cite{rasmussen2006gaussian}) so that
\begin{equation*} 
\mathbf{g} \sim \mathcal{GP}\left(0, k_{SE}(\mathbf{X},\mathbf{X'})\right)
\end{equation*}

The joint prior distribution of previous observations $\mathbf{y}$ and function values $\mathbf{g}_{*}$ evaluated at test inputs is then \cite{rasmussen2006gaussian}
\begin{align*}
\begin{bmatrix}
\mathbf{y} \\
\mathbf{g}_{*}  
\end{bmatrix}
\sim
\mathcal{N}\left(
\boldsymbol{0}, \; 
\Bigg[
\begin{matrix}
\mathbf{K}(\mathbf{X},\mathbf{X}) + \sigma_{\epsilon}^2\mathbf{I} & \mathbf{K}(\mathbf{X},\mathbf{X}_{*}) \\
\mathbf{K}(\mathbf{X}_{*},\mathbf{X}) & \mathbf{K}(\mathbf{X}_{*},\mathbf{X}_{*})
\end{matrix}\Bigg]\right)
\end{align*} 
where a subscript asterisk, such as in test inputs $\mathbf{X}_{*}$, indicates a reference to the test set quantity, and $\mathbf{g}_{*} \coloneqq \mathbf{g}\left(\mathbf{X}_{*}\right)$. We introduce the following notations for the corresponding kernel matrices, namely, $\mathbf{K}=\mathbf{K}(\mathbf{X},\mathbf{X})$, $\mathbf{K}_{*}=\mathbf{K}(\mathbf{X},\mathbf{X}_{*})$, $\mathbf{K}_{**}=\mathbf{K}(\mathbf{X}_{*},\mathbf{X}_{*})$. It follows that the joint conditional posterior distribution of $\mathbf{g}_{*}$ is also a GP, that is, $\left(\mathbf{g}_{*}| \mathbf{X},\mathbf{y},\mathbf{X}_{*}\right)\sim \mathcal{GP}\left(\boldsymbol{\mu}_{*}, \boldsymbol{\Sigma}_{*}\right)$, where \cite{rasmussen2006gaussian}:
\begin{subequations} \label{eq3:GP}
\begin{align}
\boldsymbol{\mu}_{*}&=\mathbf{K}_{*}^{\top}\left(\mathbf{K}+\sigma_{\epsilon}^2\mathbf{I}\right)^{-1}\mathbf{y} \label{eq3a:GP} \\
\boldsymbol{\Sigma}_{*}&=\mathbf{K}_{**}-\mathbf{K}_{*}^{\top}\left(\mathbf{K}+\sigma_{\epsilon}^2\mathbf{I}\right)^{-1}\mathbf{K}_{*} \label{eq3b:GP}
\end{align}
\end{subequations}

Thanks to \eqref{eq3a:GP}, we can use training data set $\mathcal{D}$ to infer the voltage sensitivity over any arbitrary test data set $\mathcal{T}$. The posterior mean \eqref{eq3a:GP} can be used directly as the best estimate for voltage sensitivity (i.e., $\boldsymbol{\hat{\alpha}}_{GP} \coloneqq \boldsymbol{\mu}_{*}$), whereas the posterior covariance \eqref{eq3b:GP} can be used to compute empirical confidence intervals that quantify estimation uncertainty. This is particularly useful for deciding whether to reassess voltage sensitivity in a certain operating region; for example, a large value of \eqref{eq3b:GP} indicates poor generalization, thereby identifying undersampled regions of the load space. By setting $\boldsymbol{\hat{\alpha}}$ in \eqref{eq6:main} according to the GP-estimate calculated by \eqref{eq3a:GP}, the proposed parameterized model achieves more accurate modeling for varied loading scenarios. Its finalized form is given by
\begin{equation}\label{eq4:GP} 
\hat{\boldsymbol{v}}=v_{0}\mathbf{1}_{n}+\mathbf{\hat{R}}\boldsymbol{p}+\mathbf{\hat{X}}\boldsymbol{q}
\end{equation}
where 
\begin{gather*}
\mathbf{\hat{R}}=\mathbf{M}^{-1}\!\left(2\mathbf{I}_{n}-\mathcal{D}\left(\mathbf{K}_{*}^{\top}\left(\mathbf{K}+\sigma_{\epsilon}^2\mathbf{I}\right)^{-1}\mathbf{y}\right)\right)\mathcal{D}\left(\boldsymbol{r}\right)\mathbf{M}^{-\top},\\
\mathbf{\hat{X}}=\mathbf{M}^{-1}\!\left(2\mathbf{I}_{n}-\mathcal{D}\left(\mathbf{K}_{*}^{\top}\left(\mathbf{K}+\sigma_{\epsilon}^2\mathbf{I}\right)^{-1}\mathbf{y}\right)\right)\mathcal{D}\left(\boldsymbol{x}\right)\mathbf{M}^{-\top}.
\end{gather*}

The above expressions for $\mathbf{\hat{R}}$ and $\mathbf{\hat{X}}$ clearly show that a sufficient condition for model linearity is the assumption that the test inputs (pertaining to the load injection points) are given. This assumption holds because the net injected powers are considered as known inputs for the power flow analysis. That is, the linear approximation of the proposed parameterization is valid within a predefined input domain jointly determined by the sets $\mathcal{D}$ and $\mathcal{T}$, which renders \eqref{eq4:GP} linear. Importantly, the use of GPs allows us to continuously update (i.e., tune) $\boldsymbol{\hat{\alpha}}$ and thus $\mathbf{\hat{R}}$ and $\mathbf{\hat{X}}$ in \eqref{eq4:GP}, concurrently with a new (single) test instance $\mathbf{x}_{*}$, or periodically in a certain load region (multiple test points), which allows great flexibility and better accuracy (as demonstrated below). 
\begin{remark}
The resulting model, once parameterized as previously described, is essentially a single iteration of \eqref{eq4:GP}. As such, the proffered model can be incorporated into existing numerical frameworks without any additional computational costs. However, in an optimal power flow (OPF) setting where the operating points change (and therefore cannot be assumed to be given), to preserve linearity and use \eqref{eq4:GP} as a stand-alone linearization, predefined test inputs can be used to initialize the corresponding matrices (e.g., nominal operating points can serve as an initial guess). In this case, the approximate OPF solution may need to be updated to reflect large changes in system conditions. Alternatively, an iterative, successive approximation scheme to the OPF of \cite{9683780} can be employed, where each step iteratively updates $\mathbf{\hat{R}}$ and $\mathbf{\hat{X}}$; each recalculation of the respective matrices thus results in a new set of PLPF equations. 
\end{remark}
\begin{remark}
The standard GP has been demonstrated to be a good model for the voltage sensitivity function for small- and medium-sized systems (see Section~\ref{Section V}). However, its applicability to large-scale systems comes with the potential for additional adaptations. Namely, the inversion of $\mathbf{K}+\sigma_{\epsilon}^2\mathbf{I}$ in \eqref{eq3:GP} scales with $\mathcal{O}(N^{3})$, where $N$ is the training data size. With $\left(\mathbf{K}+\sigma_{\epsilon}^2\mathbf{I}\right)^{-1}\mathbf{y}$ and $\left(\mathbf{K}+\sigma_{\epsilon}^2\mathbf{I}\right)^{-1}$ cached after training, calculating the mean in \eqref{eq3a:GP} and the covariance in \eqref{eq3b:GP} requires $\mathcal{O}(N)$ and $\mathcal{O}(N^{2})$ computations for a new observation, respectively. This practically limits direct implementation of the exact GP to systems with a number of nodes $n<1000$. Nevertheless, this is by no means discouraging because multiple approximation methods have been developed that maintain high accuracy while drastically reducing computational time (see e.g., \cite{8951257} for a detailed review on widely adopted sparse approximations). This, however, is left for future work.
\end{remark}

\subsection{Implementation Details} \label{Subsection IV-3}
In this subsection, we describe one approach to constructing a set of training data $\mathcal{D}=\{\mathbf{X}_{i},\mathbf{y}_{i}\}_{i=1}^{p}$ and, analogously, a set of testing data $\mathcal{T}=\{\mathbf{X}_{*_i},\mathbf{y}_{*_i}\}_{i=1}^{p_{*}}$ (where $p$ and $p_{*}$ are the number of training and test samples, respectively) that are crucial for the estimation accuracy of \eqref{eq3a:GP}, and thus the PLPF model \eqref{eq4:GP}. Note that the way $\mathcal{D}$ (resp. $\mathcal{T}$; see Section \ref{Subsection V-1}) is generated is not unique, rather it is user-specific. Ultimately, it makes no difference if $\mathcal{D}$ is generated differently -- the proposed parameterization procedure remains the same. The example below uses a fixed-granularity-based incremental method and only 20 data samples; this approach provides a simple and straightforward way of constructing $\mathcal{D}$. As expected, increasing the number of training data points increases the accuracy of GP-aided parameterization, but at the expense of computational costs (see Fig.~4). With only $p=5$ samples (not depicted), the mean squared error of the voltage sensitivity estimate is an order of magnitude larger ($1.1\cdot10^{-3}$) compared to the cases shown when $p \in [10,50]$.

Let us now construct a training data set $\mathcal{D}$ by solving \eqref{eq02:main} for $p$ different input vectors $\mathbf{x}_{i} \in \mathbb{R}^{n}$ (recall, $n$ is the number of non-root nodes). Here, each $\mathbf{x}_{i} \in \mathbb{R}^{n}$ collects complex net power injections pertaining to $p$ different operating states. In particular, we generate a deliberately small input sample $\{\mathbf{x}_{i}\}_{i=1}^{p=20}$, each with components pulled uniformly from the interval $[-2\boldsymbol{s}^{ref},-\boldsymbol{s}^{ref}] \cup [\boldsymbol{s}^{ref},2\boldsymbol{s}^{ref}]$, where $\boldsymbol{s}^{ref} \in \mathbb{R}^{n}$ is the vector of original complex power injections in the observed test system which serves as a reference. Thereafter, the associated response variables $\mathbf{y}_{i}=g(\mathbf{x}_{i}) \in \mathbb{R}^{n}$ are calculated using \eqref{eq7:main} for each obtained noise-free\footnote{To ensure a fair comparison later in the numerical analysis, we assume noise-free observations and thus omit the error variance in \eqref{eq4:GP} by setting $\sigma_{\epsilon}^2=0$.} $\mathbf{x}_{i}$, as previously described. The column vector inputs for all $p$ samples are aggregated in the $(n \cdot p) \times m$ matrix $\mathbf{X}$, while the column vector outputs are collected in the $(n \cdot p) \times 1$ vector $\mathbf{y}$, thereby obtaining $\mathcal{D}=\left(\mathbf{X},\mathbf{y}\right)$ where $|\mathcal{D}|=n \cdot 20$. Note that for the \textit{i}th sample, $\mathbf{X}_{i} = [\mathbf{x}_{i}^{re},\mathbf{x}_{i}^{im}] \in \mathbb{R}^{n \times m}$, $m=2$ is due to the decomposition of $\mathbf{x}_{i}$ into its real and imaginary parts, that is, $\boldsymbol{p}$ and $\boldsymbol{q}$, respectively.

\begin{figure}[!t]
\centering
\includegraphics[height=4.5cm]{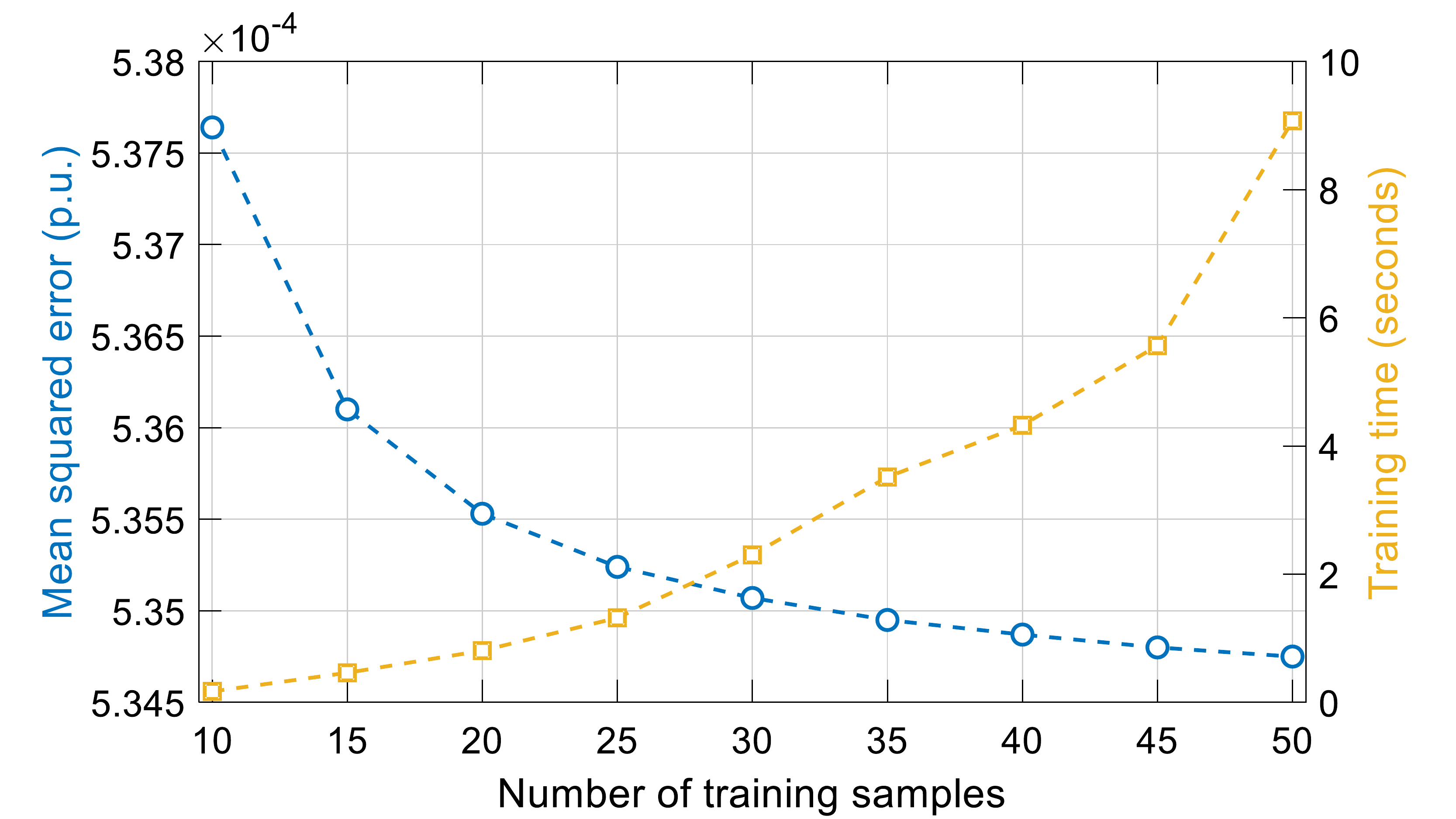}
\caption{The mean squared error of the voltage sensitivity estimates at new (test) operating point (left y-axis) and the corresponding GP training time (right y-axis) as a function of the number of training samples $p \in [10,50]$ for IEEE 33-bus system.}
\label{4}
\end{figure}

From the computational point of view, the described parameterization is simple and inexpensive (see Section \ref{Subsection V-6}). It can be summarized as follows:
\begin{list}{}{}
    \item \textit{Step 1:} Create $p$ input-output pairs needed for training ($p$ can be very small as GP works well on small data sets). Inputs $\mathbf{X} \in \mathbb{R}^{(n \cdot p) \times 2}$ are generated by changing the reference active and reactive load powers of examined test cases by at most $+200\% \; \left(-300\% \right)$ with equally-spaced steps. The corresponding outputs $\mathbf{y}^{(n \cdot p) \times 1}$ are calculated using \eqref{eq7:main}. This formalizes the training data set $\mathcal{D}=\left(\mathbf{X},\mathbf{y}\right)$. Similarly, create $p_{*}$ input-output testing pairs as described in Section \ref{Subsection V-1}. 
    \item \textit{Step 2:} Specify the desired GP prior for $\mathbf{g}(\cdot)$ (i.e., choose a mean and covariance function) and train GP model using previously created training data $\mathcal{D}$. The result of the training process is $\boldsymbol{\hat{\alpha}}_{GP}$ formalized by \eqref{eq3a:GP} and therefore the final model \eqref{eq4:GP}. 
\end{list}

\section{Numerical Analysis and Results} \label{Section V}
In this section, comparative case studies are conducted to demonstrate the performance of the proposed model. A total of six test systems are used, ranging in size from 22 to 141 buses with different loading conditions. The first comparative analysis, which is presented in Section \ref{Subsection V-3}, was conducted to compare different linear cold-start models\footnote{Note that we compare our model primarily with physics-based linear power flow models, which are open-source (if applicable) and therefore easy to reproduce, which is important given the numerous simulated loading scenarios. We put emphasis on models that are of similar form to the \textit{sDistFlow} model, such as our model, and therefore share similarities with our model.}, including the model proposed here, \textit{sDistFlow}, and decoupled linear power flow model \cite{7782382}. The abbreviations used for each of these models, in order of mention, are PLPF, SDF, and DLPF. A comparison with the \textit{Lossy DistFlow} model of \cite{8907467} is presented separately and in more detail in Section \ref{Subsection V-4}. Using variationally sparse methods similar to the previous remark (see \textit{Remark 4}), the results presented in this section can be adapted to larger systems in a straight forward manner. 

\subsection{Case Study} \label{Subsection V-1}
The model is evaluated numerically using the IEEE 33-bus, 69-bus, and 123-bus radial test systems, the last of which was modified for single-phase analysis (see \cite{7031975} for more details). In addition to the above IEEE test systems, test cases of radial distribution systems with 22-bus, 85-bus, and 141-bus available in \textsc{Matpower} \cite{5491276} are also used. The benchmark solutions, generated by \textsc{Matpower} (Newton-Rapson power solver \cite{5491276}), are regarded as \textit{ground truth} voltages. To emulate varying operational conditions and illustrate that our model is applicable to larger voltage variations, we use (\textit{i}) numerical continuation, and (\textit{ii}) Monte Carlo methods. In the former case, the reference (base) loads were scaled by a factor $k$, ranging from -2 to 2 with a granularity of 0.0345, thus constructing a total of 30 test samples. In the latter case, 10,000 different random loading scenarios are simulated.

\subsection{Evaluation Metrics} \label{Subsection V-2}
For each distribution test system, we record the approximation errors of linear power flow models by comparing their approximate voltage solutions against the true values from the exact nonlinear power flow model (\textsc{Matpower}). To this end, we use the maximum and mean estimation errors between the vector of exact voltage magnitude solution $\boldsymbol{V}$ and the vector of approximate solution $\hat{\boldsymbol{V}}^{[\textit{model}]}$ determined by one of the three models, in p.u. unit system. The errors are calculated as:
\begin{subequations}\label{eq1:eval}
\begin{align}
\varepsilon_{max}^{[\textit{model}]} &= \|\boldsymbol{V} - \hat{\boldsymbol{V}}^{[\textit{model}]}\|_{\infty} \label{eq1a:eval}\\
\varepsilon_{avg}^{[\textit{model}]} &= \frac{1}{n \cdot p_{*}} \|\boldsymbol{V} - \hat{\boldsymbol{V}}^{[\textit{model}]}\|_{1} \label{eq1b:eval}
\end{align}
\end{subequations}
where $p_{*}$ is the number of testing samples (varying consumption scenarios), $n$ is the total number of non-root nodes in a distribution test system; $\|\cdot\|_{1}$ denotes the $L_{1}$--norm, and $\|\cdot\|_{\infty}$ denotes the $L_{\infty}$--norm, previously defined in Section \ref{Subsection II-1}. 

\subsection{Results of Comparative Analysis} \label{Subsection V-3}
Here, we present the comparative analysis of the aforementioned models for both lightly and heavily loaded systems, the latter of which include cases of high consumption (power absorbed from the grid, $k>0$) and high renewable generation (power injected into the grid, $k<0$). Therefore, both low-voltage and high-voltage system conditions are simulated. This allows us to asses the robustness of the proposed model. Performance indicators for base load ($k=1$) or individual cases of high load ($|k|>1$) are calculated using \eqref{eq1:eval}, where $p_{*}=1$; in all other cases, $p_{*}=30$. For brevity, we restrict the display of voltage estimates along the buses to only the IEEE 33-bus system.

\begin{figure*}[t!]
\centering
\includegraphics[width=.33\textwidth]{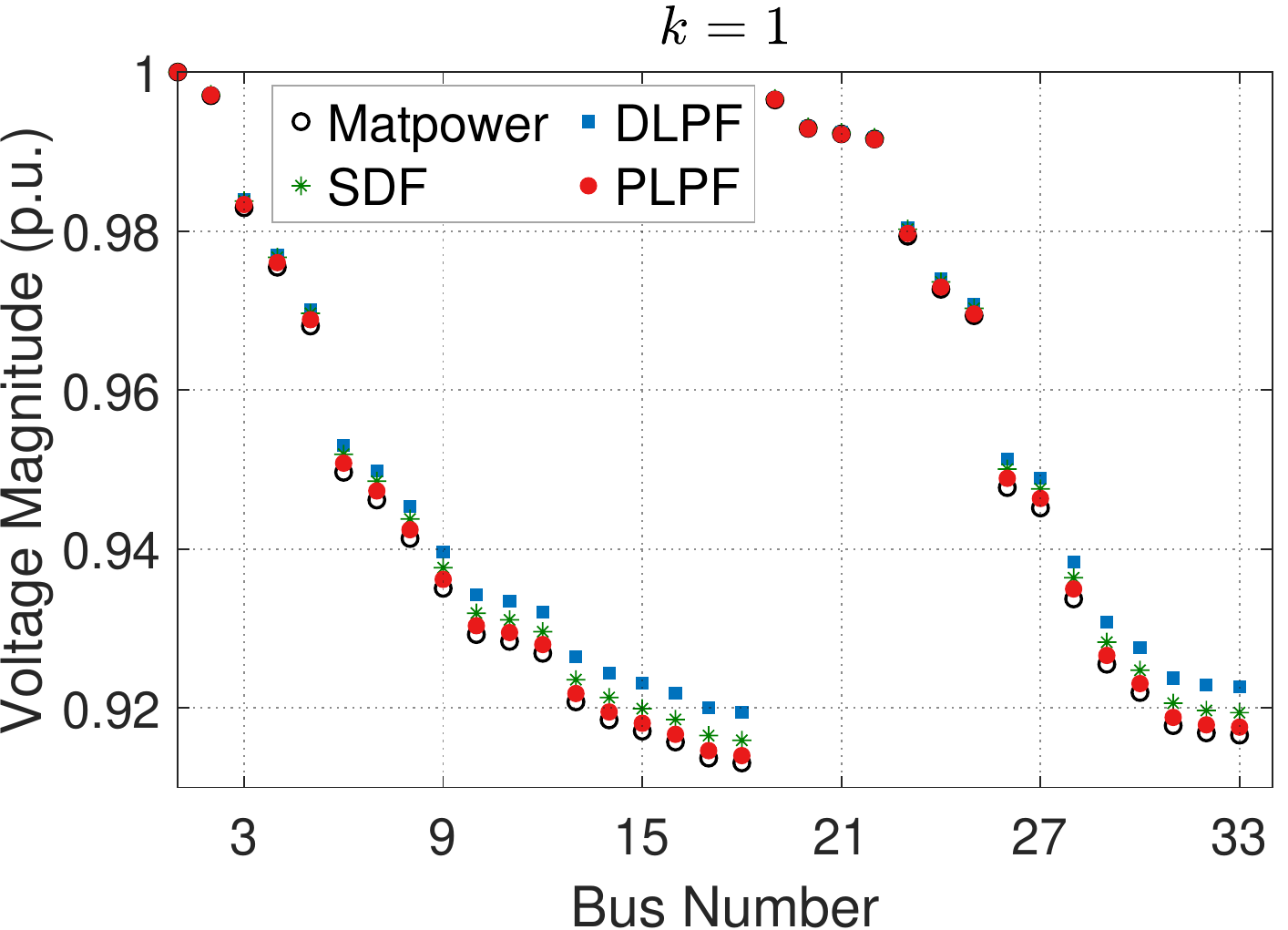}\hfill
\includegraphics[width=.33\textwidth]{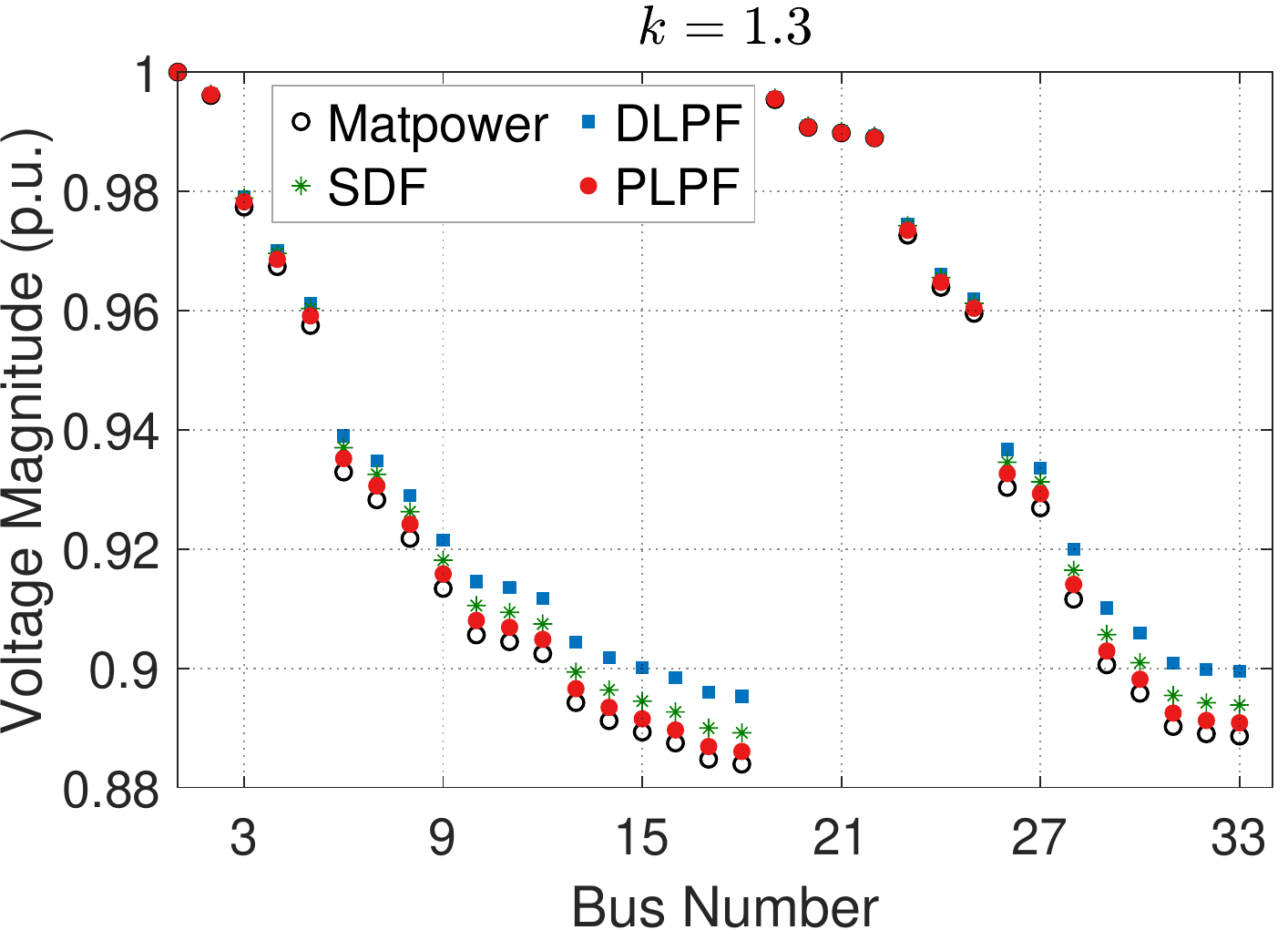}\hfill
\includegraphics[width=.33\textwidth]{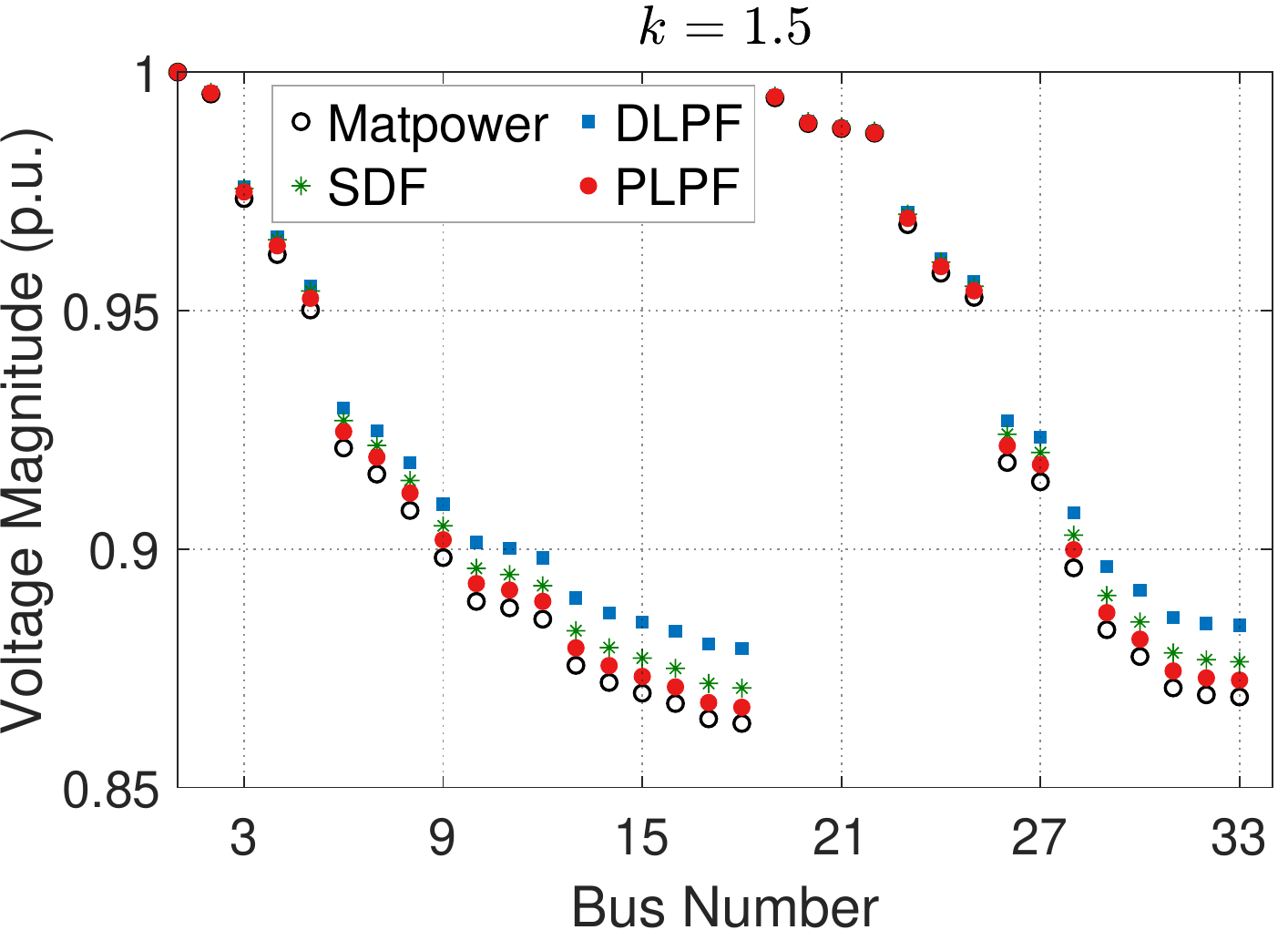}
\caption{Voltage profiles for the IEEE 33-bus system for different $k>0$ (from left to right, $k = 1$, $k = 1.3$, and $k = 1.5$, respectively).}
\label{5}
\end{figure*}
\begin{figure*}[t!]
\centering
\includegraphics[width=.33\textwidth]{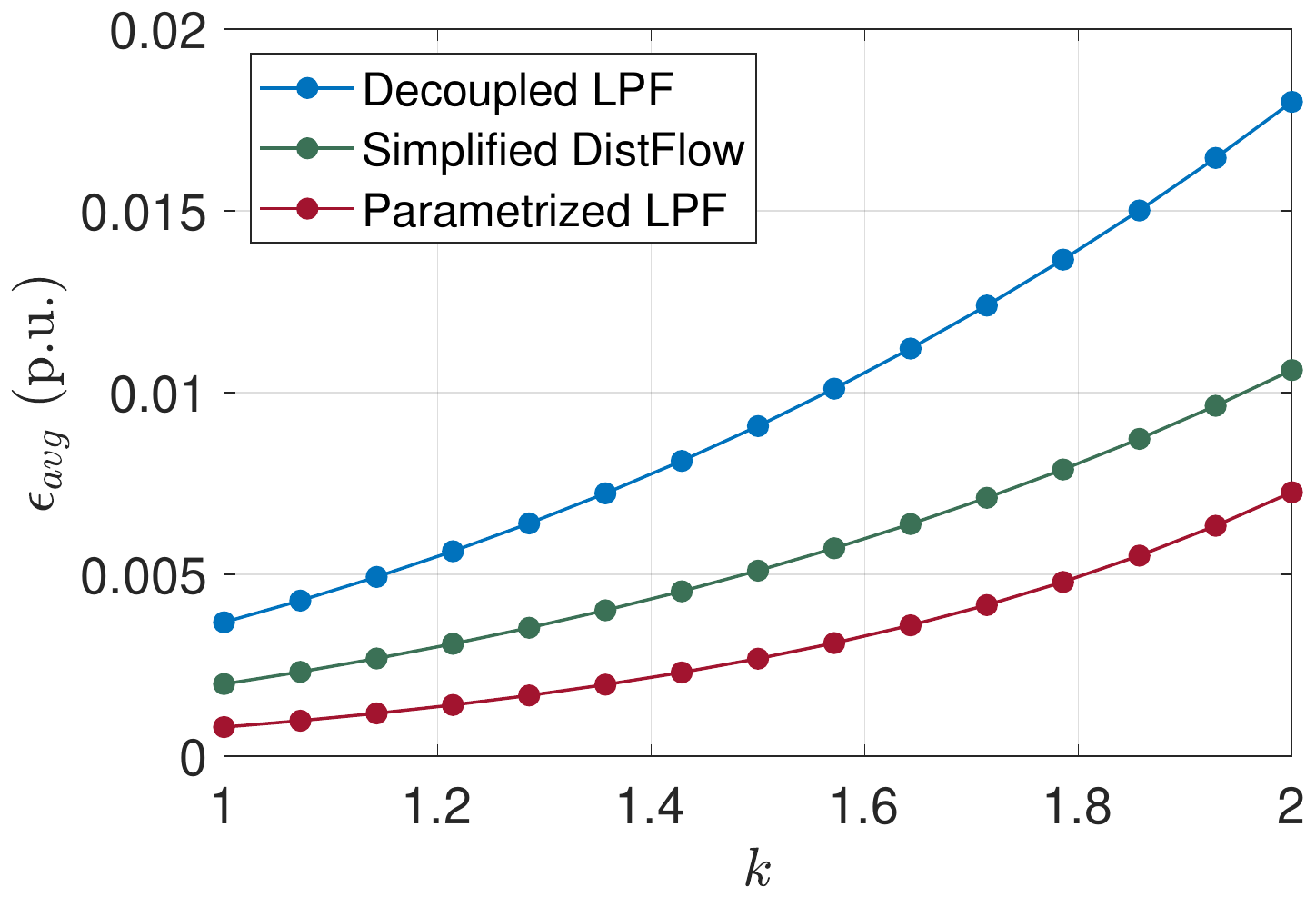}\hfill
\includegraphics[width=.33\textwidth]{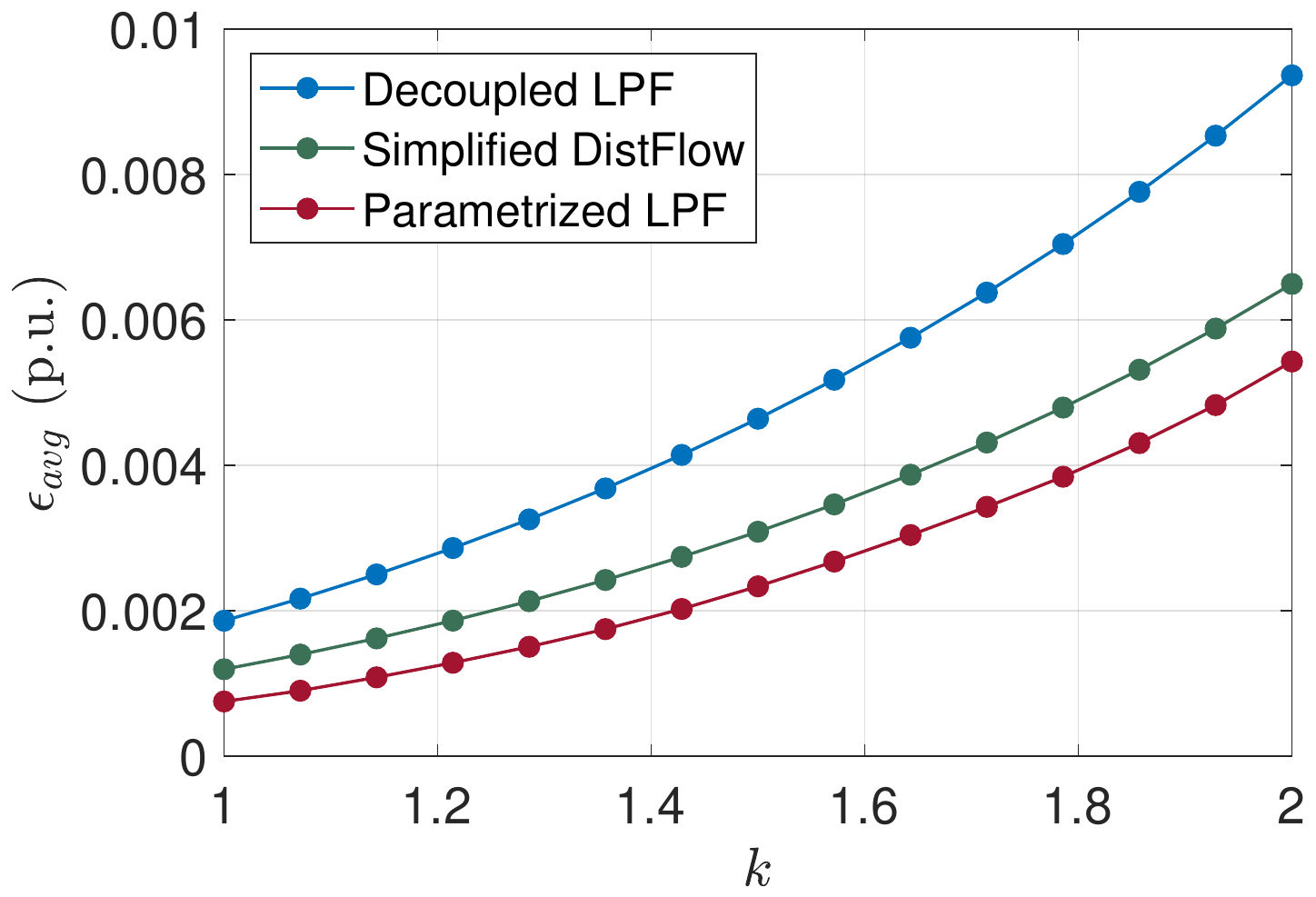}\hfill
\includegraphics[width=.33\textwidth]{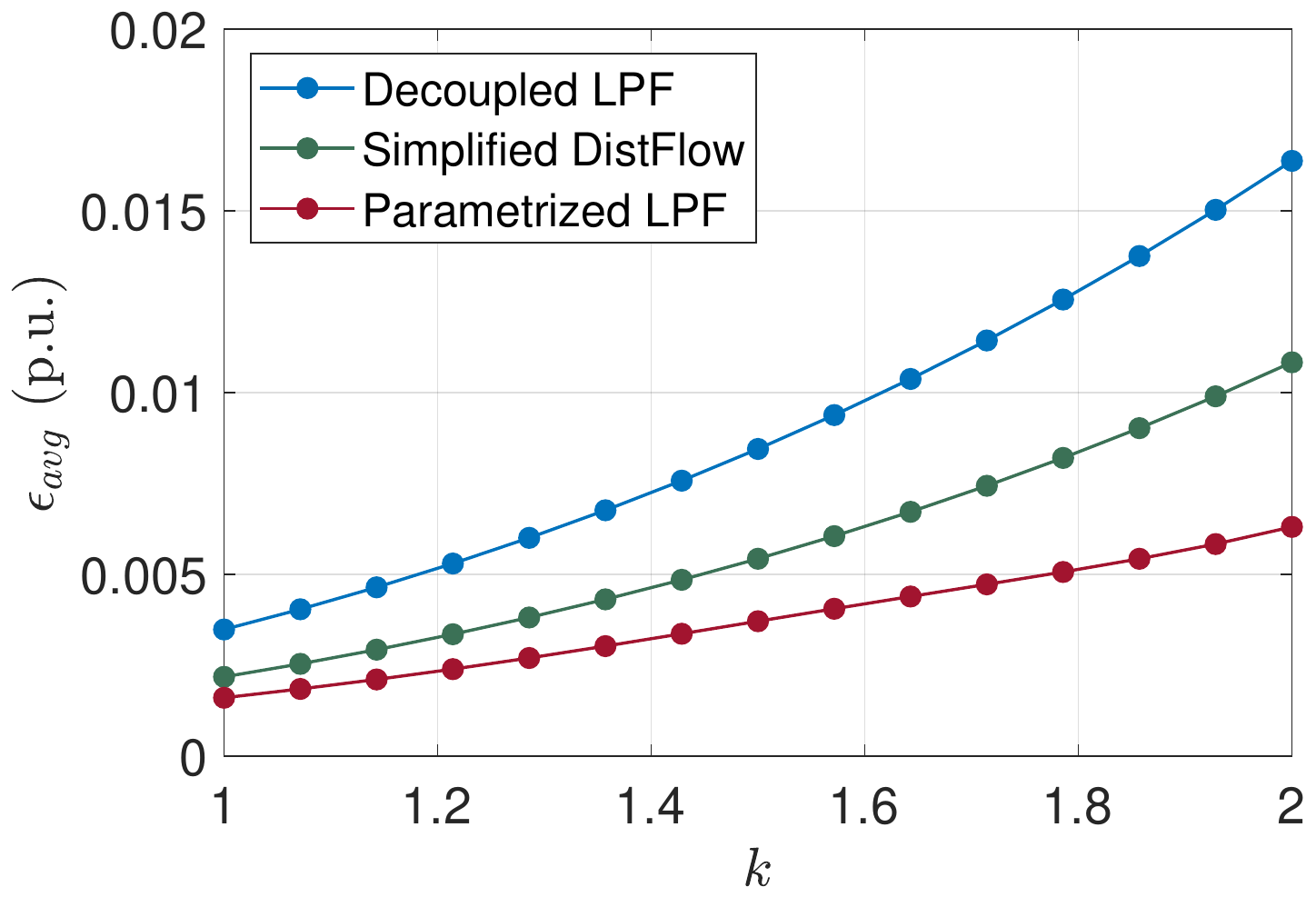}
\caption{Average estimation errors of voltage magnitudes by each of the three approximate models for different $k>0$ (from left to right, IEEE 33-bus, IEEE 69-bus, and IEEE 123-bus system, respectively).}
\label{6}
\end{figure*}

The obtained maximum and mean absolute errors (rounded to 5 decimal places) for the case of base load and cases of high load (a total of 30 test samples) are summarized in Table \ref{tab:table1} and Table \ref{tab:table2}, respectively. The results of the approximate models for the IEEE 33-bus system are shown in the left plot in Fig.~5 ($k=1$). The exact voltage solutions (obtained with \textsc{Matpower}) are plotted in black circles, while the solutions of the three approximate models are plotted as indicated in the figure legend. With a nominal feeder loading, the maximum error produced by PLPF is $1.25\cdot 10^{-3}$ (p.u.). Other results, tabulated in Table \ref{tab:table1}, confirm that our model improves the voltage estimation accuracy over the other two approximate models. 

\begin{table}[t!]
\centering
\ra{1.5}
\caption{Model Evaluation -- Base Load}
\label{tab:table1}
\resizebox{\columnwidth}{!}{%
\begin{tabular}{lcccccc}
\toprule
\multicolumn{1}{l}{\begin{tabular}[c]{@{}c@{}}Test Case\end{tabular}} & $\varepsilon_{max}^{PLPF}$ & $\varepsilon_{max}^{SDF}$ & $\varepsilon_{max}^{DLPF}$  & $\varepsilon_{avg}^{PLPF}$  & $\varepsilon_{avg}^{SDF}$  & $\varepsilon_{avg}^{DLPF}$ \\ \midrule
22-bus                                                                   & \textbf{0.00025} & 0.00030 & 0.00066 & \textbf{0.00014} & 0.00023 & 0.00040 \\
33-bus                                                                   & \textbf{0.00125} & 0.00284 & 0.00638 & \textbf{0.00080} & 0.00198 & 0.00368 \\
69-bus                                                                   & \textbf{0.00290} & 0.00388 & 0.00766 & \textbf{0.00075} & 0.00119 & 0.00186 \\
85-bus                                                                   & \textbf{0.00221} & 0.00663 & 0.01377 & \textbf{0.00180} & 0.00531 & 0.00942 \\
123-bus                                                                  & \textbf{0.00186} & 0.00255 & 0.00460 & \textbf{0.00160} & 0.00218 & 0.00348 \\
141-bus                                                                  & \textbf{0.00099} & 0.00207 & 0.00453 & \textbf{0.00071} & 0.00152 & 0.00280 \\ 
\bottomrule
\end{tabular}%
}
\end{table}
\begin{table}[t!]
\centering
\ra{1.5}
\caption{Model Evaluation -- High Load}
\label{tab:table2}
\resizebox{\columnwidth}{!}{%
\begin{tabular}{lcccccc}
\toprule
\multicolumn{1}{l}{\begin{tabular}[c]{@{}c@{}}Test Case\end{tabular}} & $\varepsilon_{max}^{PLPF}$ & $\varepsilon_{max}^{SDF}$ & $\varepsilon_{max}^{DLPF}$  & $\varepsilon_{avg}^{PLPF}$  & $\varepsilon_{avg}^{SDF}$  & $\varepsilon_{avg}^{DLPF}$ \\ \midrule
22-bus                & \textbf{0.00080} & 0.00132 & 0.00280 & \textbf{0.00020} & 0.00050 & 0.00091 \\
33-bus                & \textbf{0.01018} & 0.01573 & 0.03133 & \textbf{0.00239} & 0.00418 & 0.00795 \\
69-bus                & \textbf{0.01975} & 0.02253 & 0.03929 & \textbf{0.00199} & 0.00254 & 0.00403 \\
85-bus                & \textbf{0.03471} & 0.04700 & 0.08027 & \textbf{0.00817} & 0.01156 & 0.02055 \\
123-bus               & \textbf{0.00800} & 0.01294 & 0.02173 & \textbf{0.00301} & 0.00464 & 0.00755 \\
141-bus               & \textbf{0.00648} & 0.01070 & 0.02134 & \textbf{0.00234} & 0.00323 & 0.00608 \\ 
\bottomrule
\end{tabular}%
}
\end{table}

It can be seen that with only 20 training samples, the GP with a zero basis function and $k_{SE}$ can adequately parametrize the PLPF model resulting in highly accurate voltage solutions. As expected, the approximate solution \eqref{eq4:GP} yields less accurate predictions in high-load cases compared to the base-case loading. Across all cases, however, both the maximum error $\varepsilon_{max}^{PLPF}$ and the mean error $\varepsilon_{avg}^{PLPF}$ are quite small and consistently lower than the respective errors of the other two models. This is also confirmed in Fig.~6, where the vertical axis is the mean error $\varepsilon_{avg}$ (averaged over the total number of nodes of the observed test system) produced by each of the three models, and the horizontal axis represents the test system with different loading. The estimation errors shown in Fig.~6 correspond to 15 different positive rated load levels, $k>0$. Similar results are obtained for $k<0$ (cases of reverse power flows) and are therefore not plotted, but are included in the final error calculations presented in Table \ref{tab:table2}. As evident from the presented results, the DLPF systematically overestimates voltage magnitudes even more than the SDF; this behavior is consistent across all test cases and all loading scenarios. 

We next repeat the comparison for cases of higher feeder loading. In these cases, too, the PLPF solutions are consistently better (see middle and right plots in Fig.~5), while at heavy loading they are still more favorable compared to SDF as clearly indicated in Table \ref{tab:table2}. For demonstration, the mean prediction errors of PLPF for the three cases, where $k=\{1.3,1.5,2\}$, are lower than the respective SDF errors by a factor of 2.1, 1.9, and 1.46, respectively. However, it is clear from Fig.~6 that as the load in the test systems increases, the performance gap between the three models actually increases in favor of PLPF. This indicates that the proposed model performs better than other methods under stress conditions (i.e., higher loading). 

\subsection{Comparison with Lossy DistFlow} \label{Subsection V-4}
\begin{figure*}[t!]
\centering
\includegraphics[width=.33\textwidth]{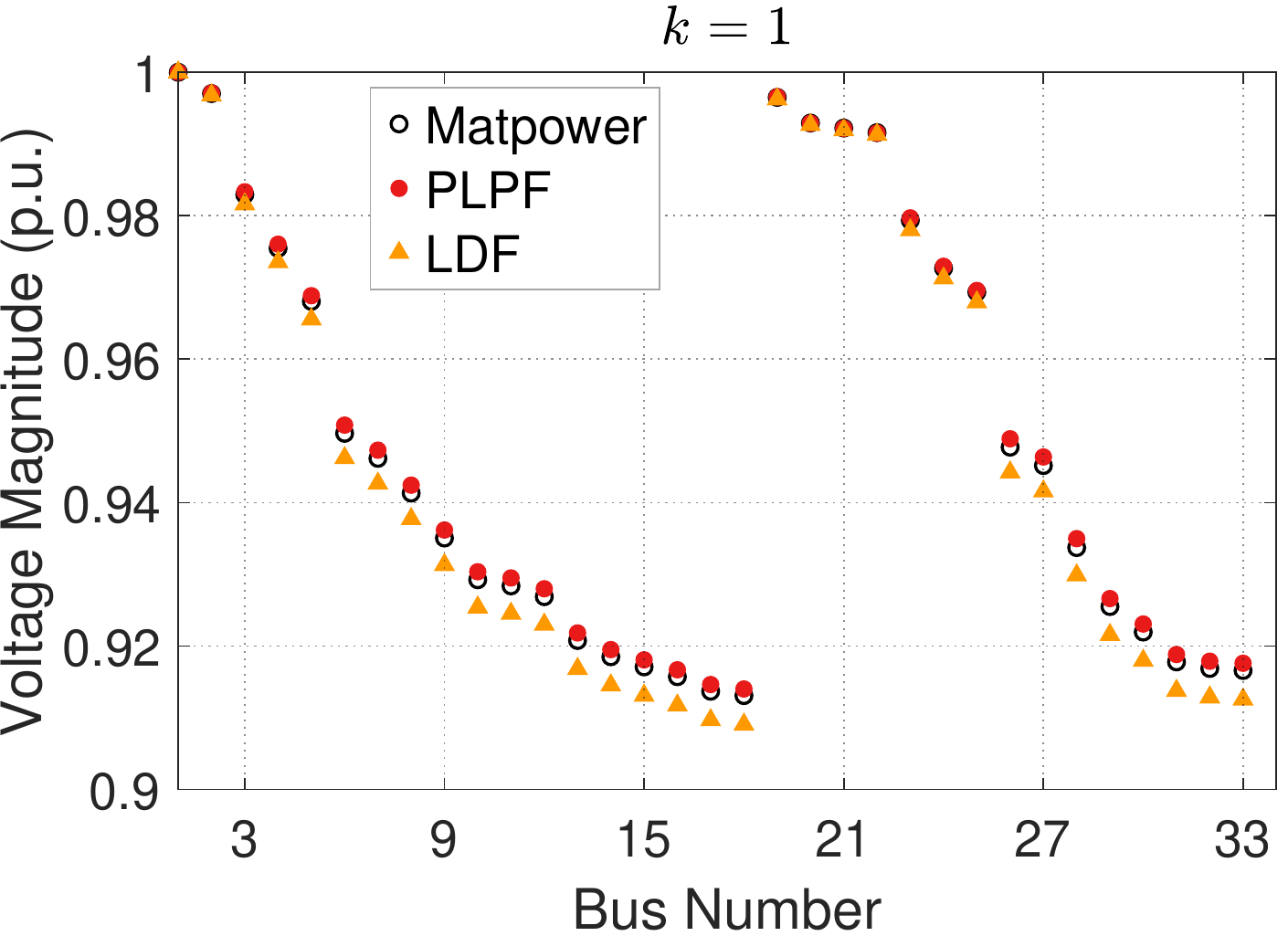}\hfill
\includegraphics[width=.33\textwidth]{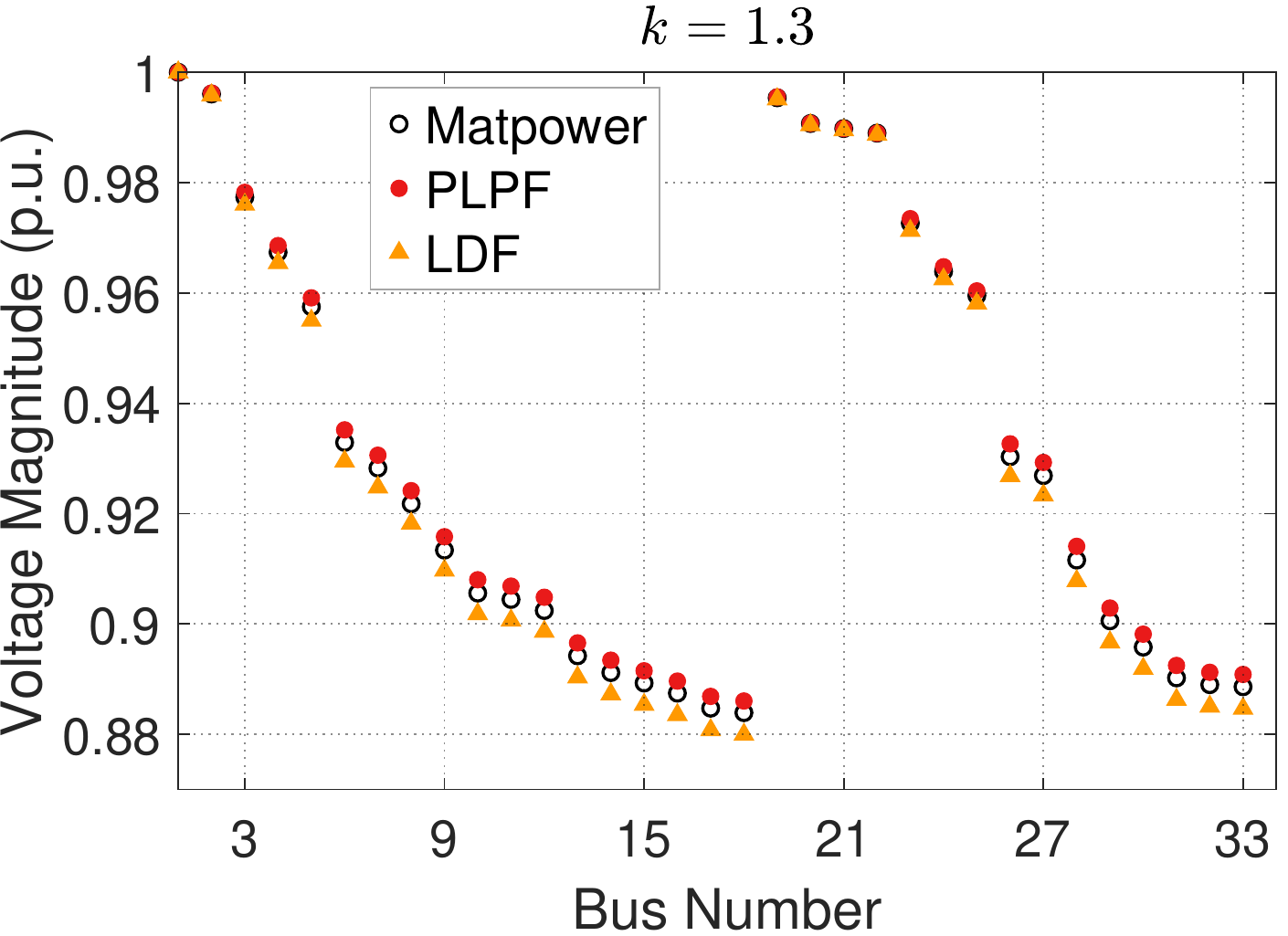}\hfill
\includegraphics[width=.33\textwidth]{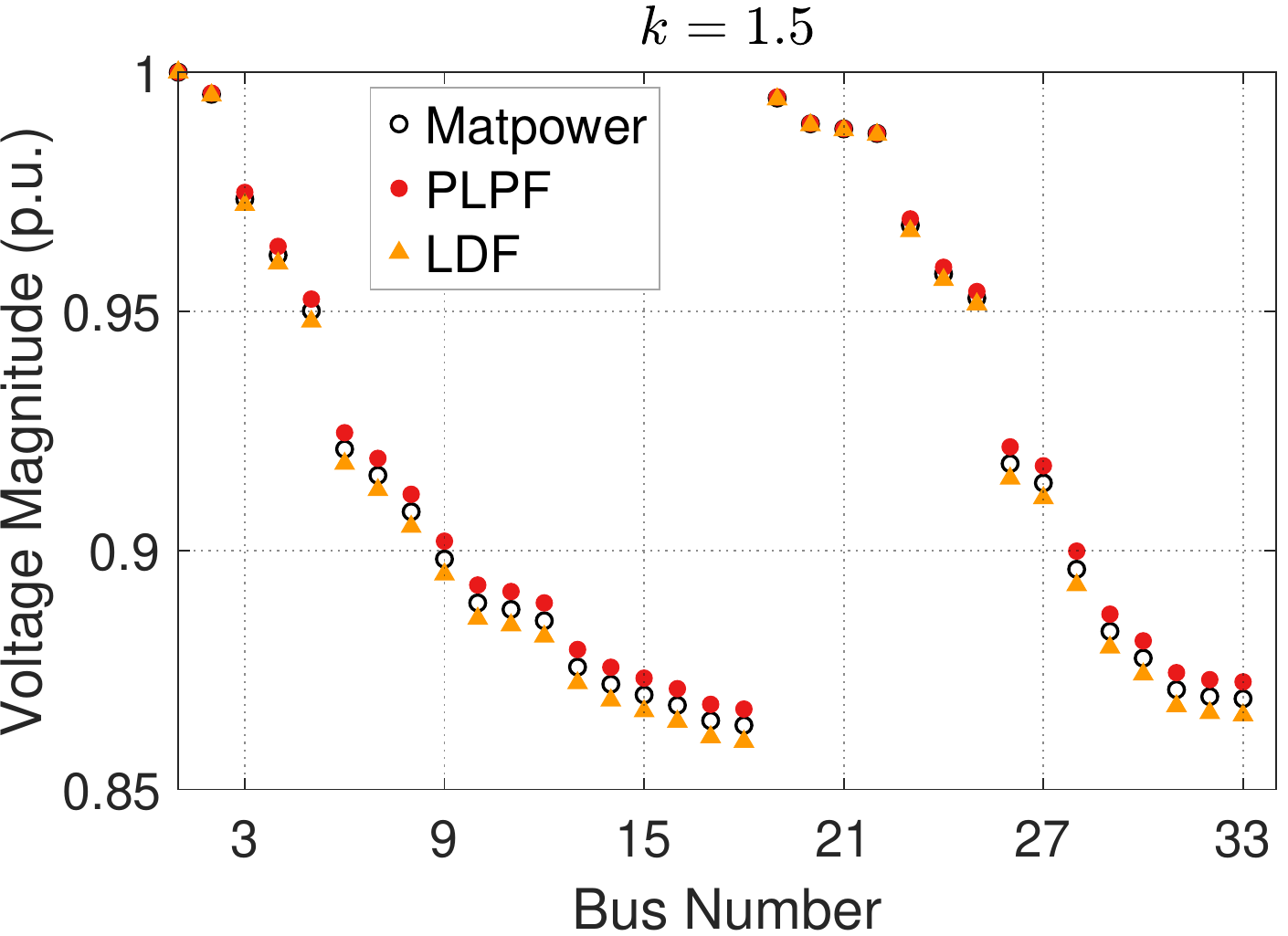}
\caption{Voltage profiles for the IEEE 33-bus system for different $k>0$ (from left to right, $k = 1$, $k = 1.3$, and $k = 1.5$, respectively).}
\label{7}
\end{figure*}
\begin{figure*}[t!]
\centering
\includegraphics[width=.25\textwidth]{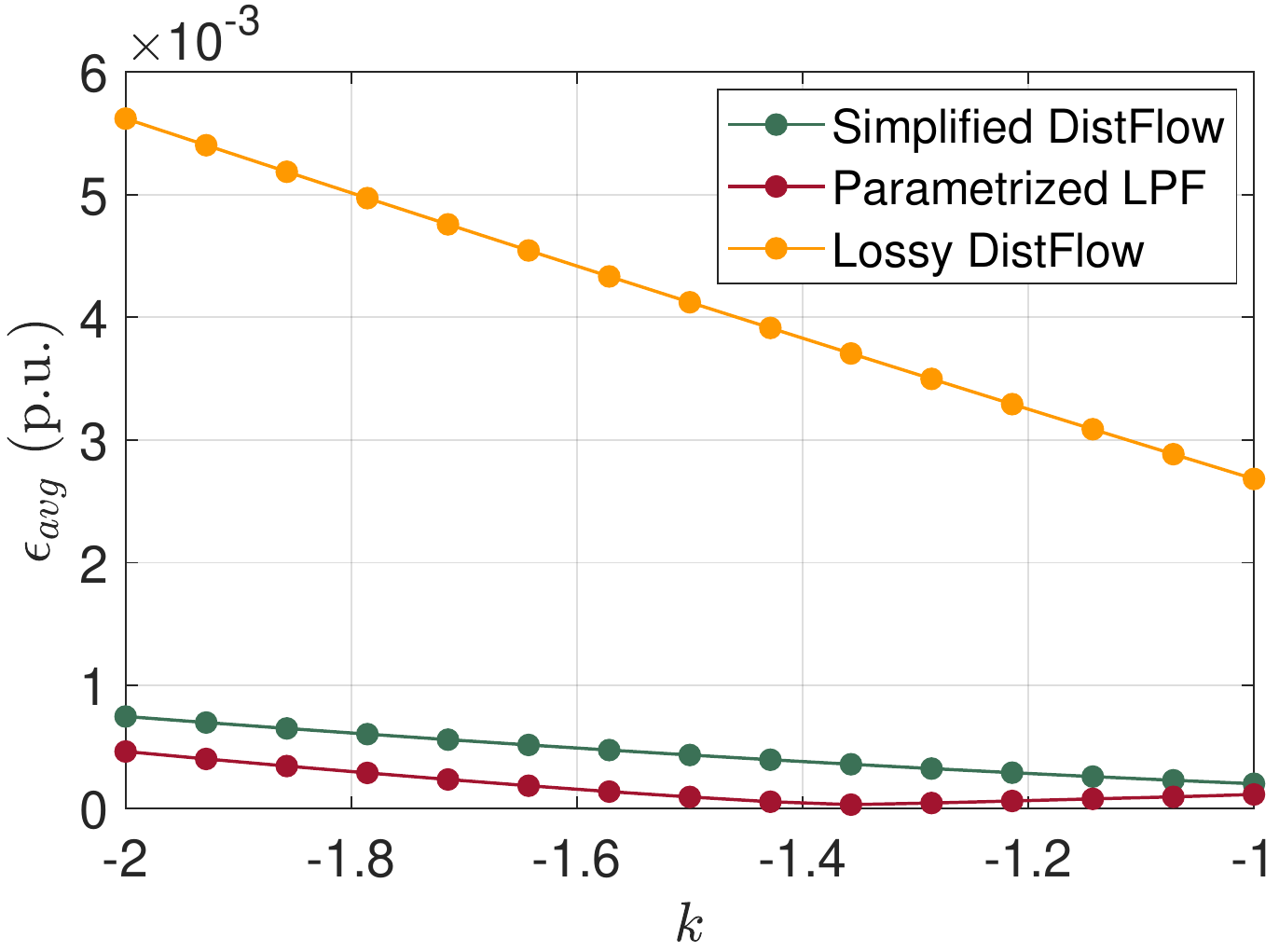}\hfill
\includegraphics[width=.245\textwidth]{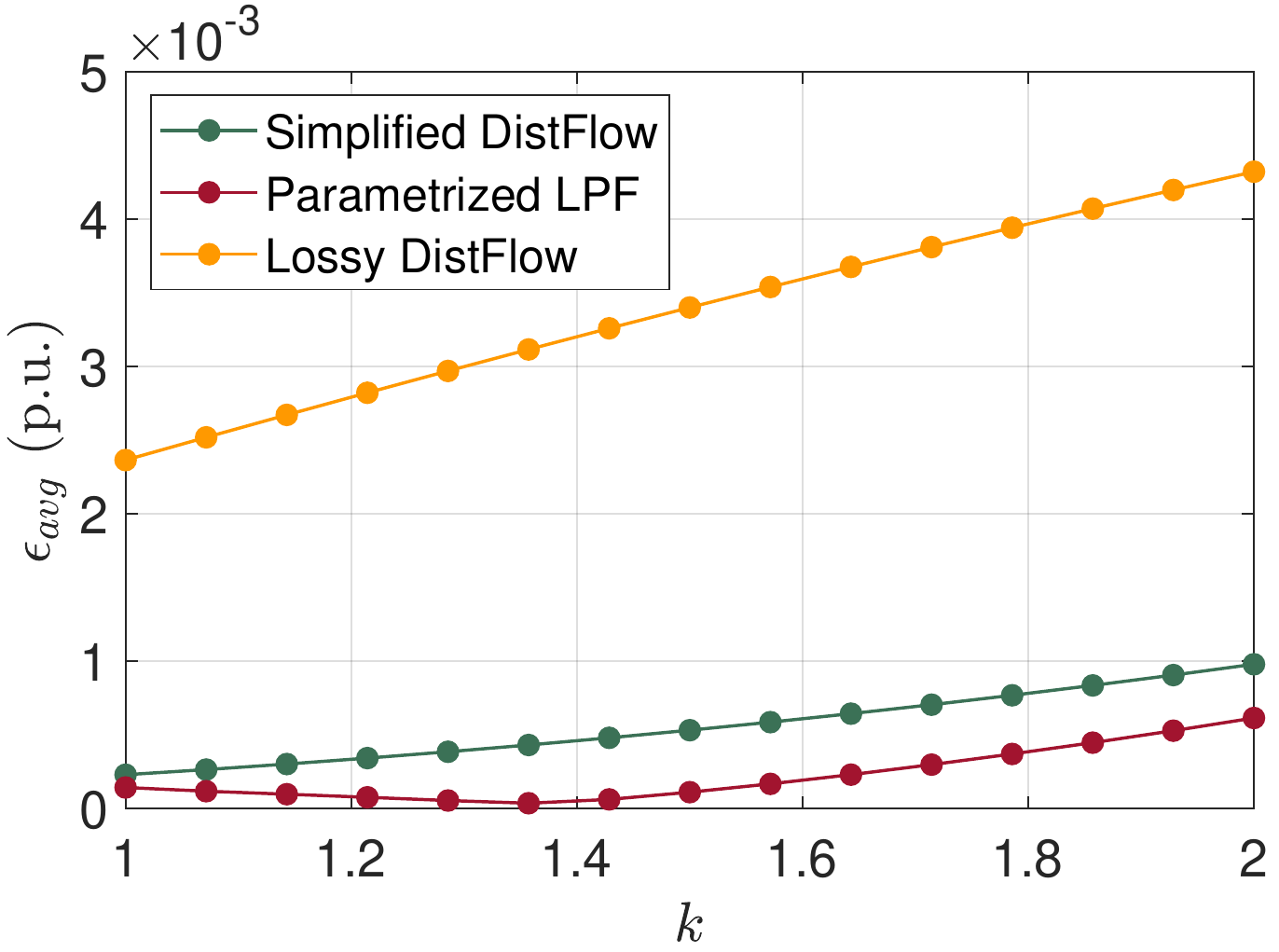}\hfill
\includegraphics[width=.255\textwidth]{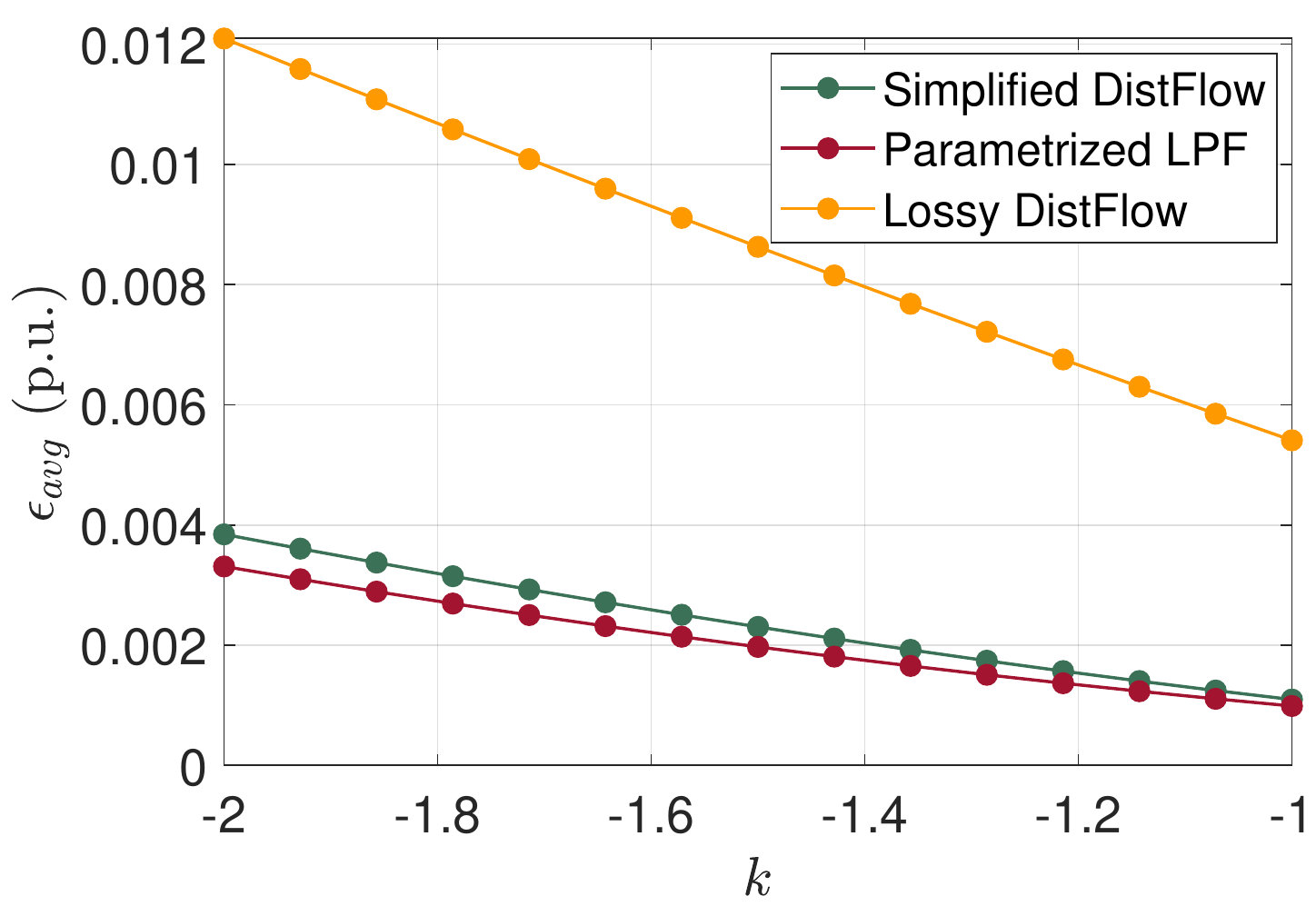}\hfill
\includegraphics[width=.25\textwidth]{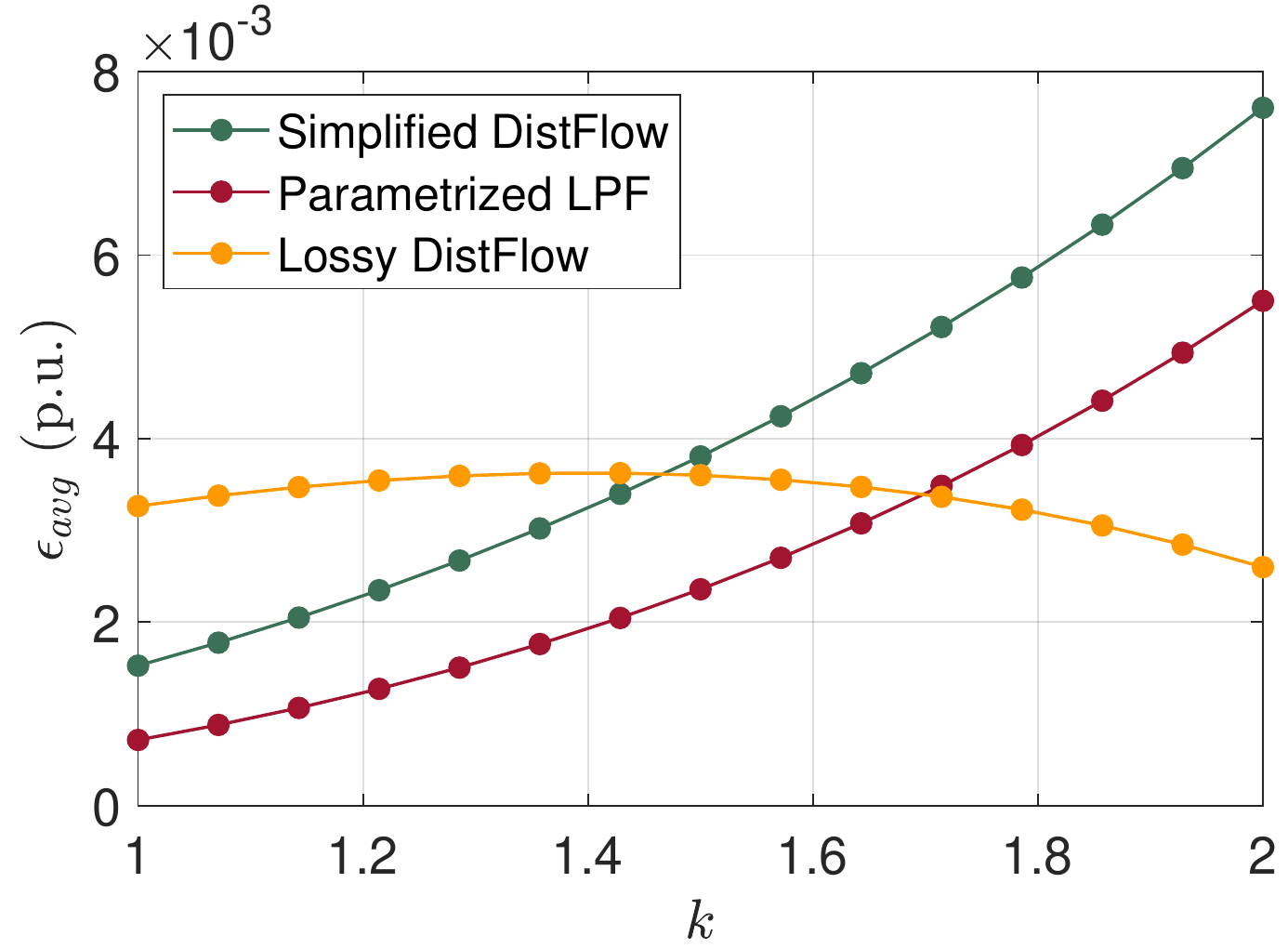}
\caption{Average estimation errors of voltage magnitudes by each of the three approximate models for different $k$ (from left to right, 22-bus system ($k<0$), 22-bus system ($k>0$), 141-bus system ($k<0$), and 141-bus system ($k>0$) respectively).}
\label{8}
\end{figure*}

\begin{table}[t!]
\centering
\ra{1.5}
\caption{Comparison with Lossy DistFlow \cite{8907467} - Base Load}
\label{tab:table3}
\resizebox{\columnwidth}{!}{%
\begin{tabular}{lcccccc}
\toprule 
\multicolumn{1}{l}{\begin{tabular}[c]{@{}c@{}}Test Case\end{tabular}} & 22-bus & 33-bus & 69-bus & 85-bus & 123-bus & 141-bus \\ \midrule
$\varepsilon_{avg}^{PLPF}$ & \textbf{0.00014} & \textbf{0.00080} & \textbf{0.00075} & \textbf{0.00180} & \textbf{0.00160} & \textbf{0.00071} \\ 
$\varepsilon_{avg}^{LDF}$ & 0.00236 & 0.00288 & 0.00112 & 0.00206 & 0.00494 & 0.00326 \\ \midrule
$\varepsilon_{max}^{PLPF}$ & \textbf{0.00025} & \textbf{0.00125} & \textbf{0.00290} & \textbf{0.00221} & \textbf{0.00186} & \textbf{0.00099}  \\ 
$\varepsilon_{max}^{LDF}$ & 0.00314 & 0.00402 & 0.00327 & 0.00261 & 0.00579 & 0.00409 \\
\bottomrule
\end{tabular}%
}
\end{table}
\begin{table}[t!]
\centering
\ra{1.5}
\caption{Comparison with Lossy DistFlow \cite{8907467} - High Load}
\label{tab:table4}
\resizebox{\columnwidth}{!}{%
\begin{tabular}{lcccccc}
\toprule
\multicolumn{1}{l}{\begin{tabular}[c]{@{}c@{}}Test Case\end{tabular}} & 22-bus & 33-bus & 69-bus & 85-bus & 123-bus & 141-bus \\ \midrule
$\varepsilon_{avg}^{PLPF}$ & \textbf{0.00020} & \textbf{0.00239} & \textbf{0.00199} & \textbf{0.00817} & \textbf{0.00301} & \textbf{0.00234} \\ 
$\varepsilon_{avg}^{LDF}$ & 0.00376 & 0.00556 & 0.00278 & 0.01037 & 0.00900 & 0.00601 \\ \midrule
$\varepsilon_{max}^{PLPF}$ & \textbf{0.0008} & \textbf{0.01018} & 0.01975 & 0.03471 & \textbf{0.00800} & \textbf{0.00648}  \\ 
$\varepsilon_{max}^{LDF}$ & 0.00733 & 0.01713 & \textbf{0.01926} & \textbf{0.02487} & 0.02017 & 0.01501 \\ 
\bottomrule
\end{tabular}%
}
\end{table}

We next compare the performance of our model with the \textit{Lossy DistFlow} model of \cite{8907467}, hereinafter abbreviated LDF. Note that we have deliberately selected the most optimal among the proposed parameterization methods from \cite[Table II]{8907467}. This ensures that LDF gives the best possible results for each of the test cases. The same was done for all studied operating settings, where power injections are changed. The results, tabulated in Table \ref{tab:table3} and Table \ref{tab:table4}, correspond to base feeder case ($k=1$) and high feeder loading ($|k|>1$), respectively. As can be seen from Table \ref{tab:table3}, our model is consistently produces lower errors than LPF in all examined test cases with the original system load (i.e., $k=1$). This coincides with the results reported in \cite{8907467}, where it was shown that LDF yields higher errors in voltage calculation than SDF for all six test cases used. Moreover, our model shows superior performance in all cases where $k<0$, which is important to consider when analyzing distribution systems with a high share of renewable generation. While our model gives better results on average, LPF shows a slightly lower maximum error, averaged over $k$ scenarios, in the case of test systems with 69 buses and 85 buses, as reported in Table \ref{tab:table4}. 

The results in Fig.~7 corroborate that the voltage solutions obtained by PLPF are closest to the exact voltage solutions generated by \textsc{Matpower} for the IEEE 33-bus system; the same is true for other test systems for $k=1$. In cases of higher feeder loading (i.e., $|k|>1$), however, different observations can be made depending on the feeder size and the scaling factor $k$, which can be seen from Fig.~8. To facilitate the exposition of the results, only the results corresponding to the smallest and largest test systems are illustrated in this figure. To provide additional insights, the SDF results are also depicted in Fig.~8. In most of the examined cases corresponding to different $k>0$, PLPF performs better than LDF, while in some cases the results are (slightly) better in favor of LDF. Interestingly, LDF has a poorer performance on smaller test cases in general (see Fig.~8, 22-bus system), while its performance improves with system size (see Fig.~8, 141-bus system), and even exceeds our model for some positive $k$ values (e.g., $k>1.7$ in the case of the 141-bus system). On the other hand, our model significantly outperforms LDF in all cases where $k=1$ and $k<0$ (e.g., Fig.~9 shows calculated voltage magnitudes of all nodes when loads are negatively doubled, $k=-2$), which can also be verified from Fig.~8. These results confirm that our model is generally more robust to network size, but also to changing system conditions, where LDF shows inconsistencies in performance.
\begin{figure}[!t]
\centering
\includegraphics[height=4.5cm]{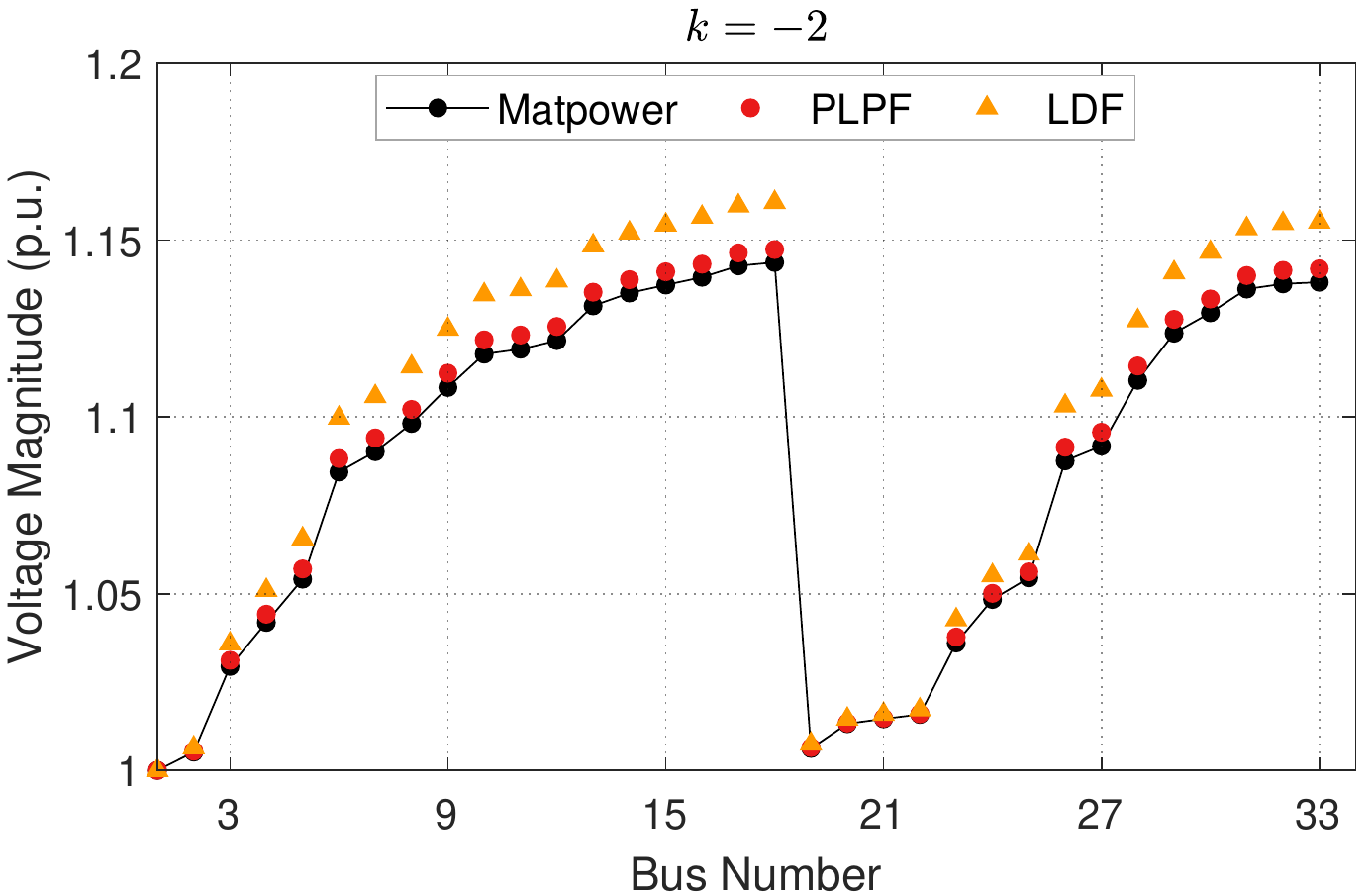}
\caption{Voltage profiles for the IEEE 33-bus system obtained by the proposed model and the model of \cite{8907467} for $k=-2$.}
\label{9}
\end{figure}

\subsection{Random Load Evaluation} \label{Subsection V-5}
For completeness, in this subsection we evaluate the proposed model under random load values, compared to the performance of SDF and LDF under the same conditions. The same evaluation metrics---defined in Section \ref{Subsection V-2}---are used (where $p_{*}=10,000$). Here, random load values are drawn from a uniform distribution in the interval $(1.5\boldsymbol{s}^{ref},0)$, where $\boldsymbol{s}^{ref}$ is the reference (base) load, previously defined. The corresponding maximum and mean values of obtained errors of the three models are summarized in Table \ref{tab:table5}. As shown in Table \ref{tab:table5}, the mean and maximum errors of PLPF are again smaller than those of the other two models.

In summary, the above numerical analyzes demonstrate that our model generally yields better accuracy than the models reported in \cite{19266,8907467,7782382} under different loading scenarios. 

\begin{table}[t!]
\centering
\ra{1.5}
\caption{Model Evaluation - Random Loads}
\label{tab:table5}
\resizebox{\columnwidth}{!}{%
\begin{tabular}{ccccccc}
\toprule
\multicolumn{1}{c}{\multirow{2}{*}{Test Case}} & \multicolumn{2}{c}{PLPF} & \multicolumn{2}{c}{SDF} & \multicolumn{2}{c}{LDF} \\ \cmidrule{2-7} 
\multicolumn{1}{c}{}    & $\varepsilon_{avg}$ & $\varepsilon_{max}$ & $\varepsilon_{avg}$ & $\varepsilon_{max}$ & $\varepsilon_{avg}$ & $\varepsilon_{max}$ \\ \midrule
22-bus                  & \textbf{0.00010} & \textbf{0.00057} & 0.00024 & 0.00090 & 0.00220 & 0.00477 \\ 
33-bus                  & \textbf{0.00051} & \textbf{0.00312} & 0.00114 & 0.00443 & 0.00262 & 0.00662 \\ 
69-bus                  & \textbf{0.00051} & \textbf{0.00816} & 0.00077 & 0.00918 & 0.00101 & 0.00476 \\ 
85-bus                  & \textbf{0.00056} & \textbf{0.00321} & 0.00278 & 0.00735 & 0.00261 & 0.00347 \\ 
123-bus                 & \textbf{0.00080} & \textbf{0.00227} & 0.00120 & 0.00300 & 0.00392 & 0.00624 \\ 
141-bus                 & \textbf{0.00035} & \textbf{0.00108} & 0.00084 & 0.00241 & 0.00273 & 0.00516 \\ 
\bottomrule
\end{tabular}
}
\end{table}

\subsection{Computational Complexity} \label{Subsection V-6}
Finally, we analyze the computational complexity of the proposed GP-based parameterization and then the PLPF model. We measure CPU time\footnote{As pointed out in \cite{8907467}, reporting time is not reliable, making it difficult to make a fair comparison with a compiled solver.} using MATLAB on a computer with the processor Intel(R) Core(TM) i5-8350U CPU @ 1.70 GHz with 8.00 GB RAM, and collect the results in Table \ref{tab:table6}. Specifically, we report running times of: 1) GP training using $p=20$ samples ($t_{train}$); 2) GP-aided parameterization for $p_{*}=30$ samples ($t_{test}$); and 3) voltage solution calculation via the PLPF model for base-case loading ($t_{PLPF}$); reported times are measured in seconds. 
\begin{table}[t!]
\centering
\ra{1.5}
\caption{CPU time in seconds}
\label{tab:table6}
\resizebox{\columnwidth}{!}{%
\begin{tabular}{lcccccc}
\toprule 
\multicolumn{1}{l}{\begin{tabular}[c]{@{}c@{}}Test Case\end{tabular}} & 22-bus & 33-bus & 69-bus & 85-bus & 123-bus & 141-bus \\ \midrule
$t_{train}$ & 0.613 & 0.722 & 3.988 & 3.284 & 2.706 & 11.888 \\ 
$t_{test}$ & 0.012 & 0.043 & 0.121 & 0.113 & 0.066 & 0.392 \\ 
$t_{PLPF}$ & $8.3\cdot10^{-3}$ & $6.0\cdot10^{-3}$ & $3.3\cdot10^{-3}$ & $3.6\cdot10^{-3}$ & $5.0\cdot10^{-3}$ & $2.0\cdot10^{-2}$ \\ 
\bottomrule
\end{tabular}%
}
\end{table}

\section{Conclusion} \label{Section VI}
The parameterized linear power flow model proposed in this paper has been shown to deliver more accurate voltage solutions compared to the \textit{simplified DistFlow} model (\textit{sDistFlow}), the most commonly used model in the relevant literature. To cover a wider range of operating points, where the performance of the \textit{sDistFlow} (but also other conventional linear power flow models) deteriorates, the proposed model is parameterized using a Gaussian Process (GP), where a small number of data samples were intentionally used in its training. Relying on a minimum of training data, realized kernel-based learning is computationally efficient. As a salient advantage, the parameterized linear power flow model shows relatively high accuracy even in stressed system conditions i.e., cases of heavy load or high renewable generation. Numerical testing indicates that the proffered model estimates voltages for the base feeder case and higher feeder loading more accurately than other linear models used for comparison. 

Given the importance of voltage solution quality in linear power flow models, the proposed model is worth further research. Potential areas of future research include: 1) using the parameterized linear power flow model as a starting point for solving relevant optimization problems for distribution system operations and analysis; 2) hyperparameter tuning (preliminary results show even better performance of the realized model with optimized hyperparameters); 3) extension to an \textit{online} setting (see \cite{8951257} for readily available extensions) -- for example, by implementing sequential re-parameterization of the model by making use of simple resetting strategies.

{\appendices
\section*{Appendix A}
C.f. \eqref{eq4} and \eqref{eq4:main}. It is clear that
\begin{equation} \label{eq1:appA}
\alpha_{ij} = \delta_{ij}\frac{\mathcal{R}_{e}\{\Delta V_{ij}\}}{\mathcal{I}_{m}\{\Delta V_{ij}\}}, 
\end{equation}
where 
\begin{gather*}
\mathcal{R}_{e}\{\Delta V_{ij}\} = r_{ij}P_{ij} + x_{ij}Q_{ij}, \\
\mathcal{I}_{m}\{\Delta V_{ij}\} = x_{ij}P_{ij} - r_{ij}Q_{ij}.
\end{gather*}
Substitute the following branch flow equations (with simplified trigonometric terms already included i.e., $\cos{\delta_{ij}} \approx 1$, $\sin{\delta_{ij}} \approx \delta_{ij}$)
\begin{gather*}
P_{ij} = \frac{1}{|z_{ij}|^2}\left(r_{ij}\left(|V_{i}|^{2}-|V_{i}||V_{j}|\right)+x_{ij}|V_{i}||V_{j}|\delta_{ij}\right)\\
Q_{ij} = \frac{1}{|z_{ij}|^2}\left(x_{ij}\left(|V_{i}|^{2}-|V_{i}||V_{j}|\right)+r_{ij}|V_{i}||V_{j}|\delta_{ij}\right)
\end{gather*}
into \eqref{eq1:appA}. After shortening and regrouping identical terms, \eqref{eq1:appA} reduces to:
\begin{equation} \label{eq2:appA}
\alpha_{ij} = \frac{|V_{i}|}{|V_{j}|}-1
\end{equation}

\section*{Appendix B}
Before we briefly explain the extension of the proposed model to three-phase setting, we first introduce the notation used below.

Let $\mathcal{P}_{i} \subseteq \{a,b,c\}$ and $\mathcal{P}_{ij} \subseteq \{a,b,c\}$ denote the sets of available phases of bus $i \in \mathcal{N}$ and line $\left(i,j\right) \in \mathcal{L}$, respectively. Hereafter, a superscript $\left(\cdot\right)^{\phi}$ is used to assign relevant electrical quantities to a specific phase \cite{7038399}. A $\sum_{i \in \mathcal{N}}|\mathcal{P}_{i}| \times 1$ vector of voltage phasors is defined as $\boldsymbol{V} \coloneqq [\boldsymbol{V}_{1}^{\top},...,\boldsymbol{V}_{n}^{\top}]^{\top}$, where $\boldsymbol{V}_{i} \coloneqq [\{V_{i}^{\phi}\}_{\phi \in \mathcal{P}_{i}}]^{\top}$ is a $|\mathcal{P}_{i}| \times 1$ vector collecting phase voltages of node $i \in \mathcal{N}$. Similarly, define vectors $\boldsymbol{v}_{i} \coloneqq [\{|V_{i}^{\phi}|^2\}_{\phi \in \mathcal{P}_{i}}]^{\top}$ and $\boldsymbol{s}_{i} \coloneqq [\{s_{i}^{\phi}\}_{\phi \in \mathcal{P}_{i}}]^{\top}$ collecting the corresponding quantities related to bus $i \in \mathcal{N}$. Likewise for lines, introduce vector $\boldsymbol{S}_{ij} \coloneqq [\{S_{ij}^{\phi}\}_{\phi \in \mathcal{P}_{ij}}]^{\top}$. Denote by $\mathbf{Z}_{ij} \in \mathbb{C}^{|\mathcal{P}_{ij}| \times |\mathcal{P}_{ij}|}$ the impedance submatrix of line $\left(i,j\right) \in \mathcal{L}$ in which all the elements other than the diagonal elements$-$self impedances $z_{ij}^{\phi\phi}-$are mutual impedances $z_{ij}^{\phi\varphi}$ ($\varphi \neq \phi$, $\phi,\varphi \in \mathcal{P}_{ij}$). As needed, a complex-valued vector (matrix) will be decomposed into its real and imaginary parts denoted as $\boldsymbol{A}^{re},\boldsymbol{A}^{im}$.

With the modeling above, the extension of (\ref{eq0:main}) to three-phase setting is given by \cite{7244261}:
\begin{equation}\label{eq9:main}
\boldsymbol{V}_{i}^{\mathcal{P}_{ij}} - \boldsymbol{V}_{j} = \mathbf{Z}_{ij}\Big[\boldsymbol{S}_{ij}^{*} \oslash \left(\boldsymbol{V}_{i}^{\mathcal{P}_{ij}}\right)^{*}\Big]
\end{equation}
which can be further reformulated as\footnote{The branch and its defining nodes do not necessarily have the same number of phases (e.g., three-phase nodes that connect to single- or two-phase laterals), hence the superscript $\left(\cdot\right)^{\mathcal{P}_{ij}}$ as in \cite{7038399}.}:
\begin{equation}\label{eq10:main}
\boldsymbol{v}_{i}^{\mathcal{P}_{ij}} - \boldsymbol{V}_{j}\odot\left(\boldsymbol{V}_{i}^{\mathcal{P}_{ij}}\right)^{*} = \mathbf{Z}_{ij}\left(\boldsymbol{P}_{ij}-j\boldsymbol{Q}_{ij}\right)
\end{equation}

Following a similar derivation for a single-phase system, a three-phase extension of (\ref{eq3:main}) can be obtained as:
\begin{equation}\label{eq12:main}
\boldsymbol{v}_{i}^{\mathcal{P}_{ij}}-\boldsymbol{v}_{j}^{\mathcal{P}_{ij}} = \left(\mathbf{I}_{|\mathcal{P}_{ij}|} + \mathcal{D}(\boldsymbol{\lambda}_{ij})\right)\big[\mathbf{Z}_{ij}\left(\boldsymbol{P}_{ij}-j\boldsymbol{Q}_{ij}\right)\big]^{re}
\end{equation}

Since the mutual impedances are much smaller than the self impedances ($z_{ij}^{\phi\varphi} \ll z_{ij}^{\phi\phi}$), they can be dropped from $\mathbf{Z}_{ij}$ as suggested in \cite{8046055}, thus leading to:
\begin{multline}\label{eq13:main}
\boldsymbol{v}_{i}^{\mathcal{P}_{ij}}-\boldsymbol{v}_{j}^{\mathcal{P}_{ij}} = \left(\mathbf{I}_{|\mathcal{P}_{ij}|} + \mathcal{D}(\boldsymbol{\lambda}_{ij})\right) \times \\ \left(\mathcal{D}(\boldsymbol{r}_{ij}^{\phi\phi})\boldsymbol{P}_{ij}+\mathcal{D}(\boldsymbol{x}_{ij}^{\phi\phi})\boldsymbol{Q}_{ij}\right)   
\end{multline}
where vector $\boldsymbol{r}_{ij}^{\phi\phi} \coloneqq [r_{ij}^{aa}, r_{ij}^{bb}, r_{ij}^{cc}]^{\top}$ (resp. $\boldsymbol{x}_{ij}^{\phi\phi} \coloneqq [x_{ij}^{aa}, x_{ij}^{bb}, x_{ij}^{cc}]^{\top}$) is obtained by $d(\mathbf{Z}_{ij}^{re})$ (resp. $d(\mathbf{Z}_{ij}^{im})$). 

Before proceeding, we first introduce a three-phase edge-to-node incidence matrix (akin to the previously defined single-phase edge-to-node incidence matrix from Section \ref{Subsection II-2}) $\bar{\mathbf{M}}=[\mathbf{M}_{0} \; \mathbf{M}]$, where $\mathbf{M}_{0}$ and $\mathbf{M}$ are respective $\sum_{\left(i,j\right) \in \mathcal{L}}|\mathcal{P}_{ij}| \times |\mathcal{P}_{0}|$ and $\sum_{\left(i,j\right) \in \mathcal{L}}|\mathcal{P}_{ij}| \times \sum_{i \in \mathcal{N}}|\mathcal{P}_{i}|$ matrices. For branch-phase index pairs $(\ell,\varphi \in \mathcal{P}_{ij})$ and bus-phase index pairs $(i,\phi \in \mathcal{P}_{i})$, the block $\bar{\mathbf{M}}(\ell,i)$ is defined as:
\begin{equation*}
    \bar{\mathbf{M}}(\ell,i) =
     \begin{cases}
      1, & \text{if} \; \ell \in \mathcal{L} \; \text{leaves} \; i \in \mathcal{N} \cup \{0\} \; \text{and} \; \phi=\varphi\\
      -1, & \text{if} \; \ell \in \mathcal{L} \; \text{enters} \; i \in \mathcal{N} \; \text{and} \; \phi=\varphi\\
      0, & \text{if} \; \ell \in \mathcal{L} \; \text{is not incident to} \; i \in \mathcal{N} \; \text{or} \; \phi \neq \varphi
    \end{cases} 
\end{equation*}

Now starting from (\ref{eq13:main}), the three-phase extension of (\ref{eq5:main}) can be derived. Its compact form is given by:
\begin{subequations} \label{eq14:main}
\begin{gather}
\begin{bmatrix}
\mathbf{M}_{0} & \mathbf{M}
\end{bmatrix}
\begin{bmatrix}
\boldsymbol{v}_{0}  \\
\boldsymbol{\hat{v}}  
\end{bmatrix}
= 
\begin{bmatrix}
\widehat{\mathbf{R}}  & \widehat{\mathbf{X}} 
\end{bmatrix}
\begin{bmatrix}
\boldsymbol{p}  \\
\boldsymbol{q}  
\end{bmatrix} \\
\widehat{\mathbf{R}} =\left(2\mathbf{I}_{|\mathcal{P}_{ij}|} - \mathcal{D}\left(\boldsymbol{\hat{\alpha}}\right)\right)\mathcal{D}(\boldsymbol{r}^{\phi\phi})\mathbf{M}^{-\top} \\ 
\widehat{\mathbf{X}} = \left(2\mathbf{I}_{|\mathcal{P}_{ij}|} - \mathcal{D}\left(\boldsymbol{\hat{\alpha}}\right)\right)\mathcal{D}(\boldsymbol{x}^{\phi\phi})\mathbf{M}^{-\top}
\end{gather}
\end{subequations}
Here, $\boldsymbol{\hat{\alpha}}$ is redefined as a $|\mathcal{P}_{ij}| \times 1$ parameter vector.}

\bibliographystyle{IEEEtran}


\newpage

 




\vfill

\end{document}